\documentclass[12pt,letterpaper]{article}          
\usepackage{osajnl}

\newcommand{\im}{{\rm i}}
\newcommand{\dif}{{\rm d}}
\newcommand{\sgn}{{\rm sgn}}
\newcommand{\point}[1]{\ensuremath{{\rm #1}}}
\newcommand{\subP}[2][P]{#2_\point{#1}}
\newcommand{\smfrac}[2]{\textstyle{\frac{#1}{#2}}}
\newcommand{\eqref}[1]{(\ref{eq:#1})}
\newcommand{\Eq}[1]{Eq.~\eqref{#1}}
\newcommand{\Fig}[1]{Fig.~\ref{fig:#1}}
\newcommand{\Section}[1]{Section~\ref{sec:#1}}

\begin{document}

\title{Morphology of the nonspherically decaying radiation beam
generated by a rotating superluminal source}

\author{Houshang Ardavan}
\address{Institute of Astronomy, University of Cambridge,
\\  Madingley Road, Cambridge CB3 0HA, UK}

\author{Arzhang Ardavan}
\address{Clarendon Laboratory, Department of Physics, University of Oxford,
\\  Parks Road, Oxford OX1 3PU, UK}

\author{John Singleton}
\address{National High Magnetic Field Laboratory, MS-E536,
\\  Los Alamos National Laboratory, Los Alamos, New Mexico 87545}
\email{jsingle@lanl.gov}

\author{Joseph Fasel and Andrea Schmidt}
\address{Process Engineering, Modeling, and Analysis, MS-F609
\\  Los Alamos National Laboratory, Los Alamos, New Mexico 87545}

\begin{abstract}
We consider the nonspherically decaying radiation field
that is generated by a polarization current 
with a superluminally rotating distribution pattern in vacuum,
a field that decays
with the distance $\subP{R}$ from its source
as $\subP{R}^{-1/2}$,
instead of $\subP{R}^{-1}$. 
It is shown 
(i) that the nonspherical decay of this emission
remains in force
at all distances from its source
independently of the frequency of the radiation,
(ii) that the part of the source
that makes the main contribution
toward the value of the nonspherically decaying field
has a filamentary structure
whose radial and azimuthal widths become narrower
(as $\subP{R}^{-2}$ and $\subP{R}^{-3}$, respectively),
the farther the observer is from the source,
(iii) that the loci
on which the waves emanating from this filament interfere constructively
delineate a radiation `subbeam'
that is nondiffracting in the polar direction,
(iv) that the cross-sectional area of each nondiffracting subbeam
increases as $\subP{R}$,
instead of $\subP{R}^2$,
so that the requirements of conservation of energy are met
by the nonspherically decaying radiation automatically,
and (v) that the overall radiation beam
within which the field decays nonspherically
consists, in general,
of the incoherent superposition of such coherent nondiffracting subbeams.
These findings are related
to the recent construction and use of superluminal sources in the laboratory
and numerical models of the emission from them.
We also briefly discuss the relevance of these results
to the giant pulses received from pulsars.
\end{abstract}


\maketitle 

\section{Introduction}
\subsection{Preamble}
Maxwell's generalization of Amp\`ere's law\cite{JacksonJD:Classical}
establishes that electromagnetic radiation can be
equally well generated
by a time-dependent electric polarization current, with a density $\partial {\bf P}/\partial t$,
as by a current of accelerated free charges with the density ${\bf j}$:
\begin{equation}
\nabla \times {\bf H} = 
\frac {4\pi}{c} {\bf j} + \frac{1}{c} \frac{\partial {\bf D}}{\partial t} =
\frac {4\pi}{c} ({\bf j}+\frac{\partial{\bf P}}{\partial t}) + \frac{1}{c} \frac{\partial {\bf E}}{\partial t};
\label{ampere}
\end{equation}
here, 
${\bf E}$ and ${\bf H}$ are the electric and magnetic fields,
${\bf D}$ is the displacement and $c$ is the speed of light {\it in vacuo}.  
A remarkable aspect of the emission from such polarization currents
is that the motion of the radiation source is not limited by $c$. 
Although the speed of charged particles cannot exceed $c$,
nothing prevents the distribution pattern of a polarization current,
created by the coordinated motion of subluminal 
particles, from moving faster than 
light.\cite{BolotovskiiBM:VaveaD,GinzburgVL:vaveaa,BolotovskiiBM:Radbcm}
Indeed, radiation from such superluminal polarization
currents has been observed in the laboratory.%
\cite{BessarabAV:FasEsi,ArdavanA:Exponr,SingletonJ:Expdes,BolotovskiiBM:Radsse}

Since electric polarization arises from separation of charges,
a polarization current is by its nature volume-distributed.
In fact, no superluminal source can be point-like; 
for, if a point source were to move faster than its own waves, 
it would generate caustics on which the field strength would diverge.\cite{BolotovskiiBM:VaveaD,ArdavanH:Genfnd}  

There is growing experimental and theoretical interest
in radiation by polarization currents
whose distribution patterns move at a superluminal speed with acceleration.\cite{BolotovskiiBM:Radsse}
One of the simplest implementations of such sources
employs distribution patterns that have the time dependence of a traveling wave
with circular superluminal motion; here, the acceleration is {\it centripetal}.
We are investigating the use of polarization currents
with such superluminally rotating distribution patterns
in applications relating to communications and radar.\cite{ArdavanA:Appgfer,ArdavanA:Exponr}
Furthermore,
one of the proposed models of the radio emission from pulsars
postulates the presence of sources of this type
in the magnetospheres of rapidly rotating neutron stars.\cite{ArdavanH:supmp,SchmidtA:Occopmlw}
The clarification of a diverse set of current questions, therefore, 
hinges on an understanding of the radiation
from superluminal polarization currents
undergoing circular motion~\cite{ArdavanH:Speapc,HannayJH:Speapc,ArdavanH:Speapc1}.

Our purposes in the present paper are
(i)~to examine the geometry of those regions within such extended sources 
that make the dominant contribution toward the radiation field 
observed at a given point and time, and 
(ii)~to identify the salient features Uof the angular distribution of 
this radiation.  
A detailed knowledge
of the extent and geometry of the contributing part of the source
is required not only for the efficient design
of practical superluminal sources of this type
({\it e.g.} for the design of the dielectric 
in which the polarization current is generated),\cite{ArdavanA:Exponr}
but also for understanding
the narrow widths of the giant pulses
that are received from pulsars.\cite{LorimerD:HbPA} 
Likewise,
a knowledge of the evolution of the angular distribution of the radiation with distance
both facilitates the experimental detection
of the tightly-beamed large-amplitude component
of the emission from such sources
and establishes a connection
between two observed features
(the nanostructure and the high brightness temperature)
of the pulsar emission.
\cite{SallmenS:Simdog,HankinsTH:Nanrbs,SoglasnovVA:GiapPB,PopovMV:GPmcrecp}

\begin{figure}
\centering
\includegraphics[width=9cm]{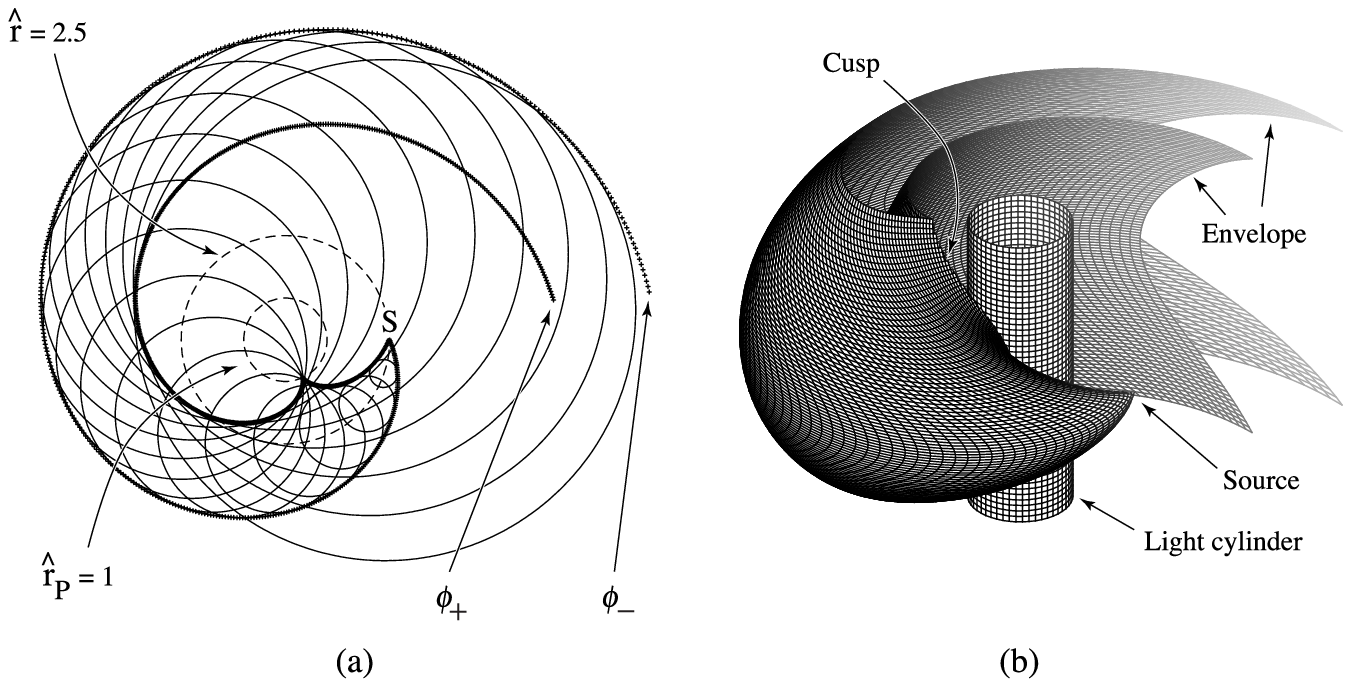}
\includegraphics[width=5cm]{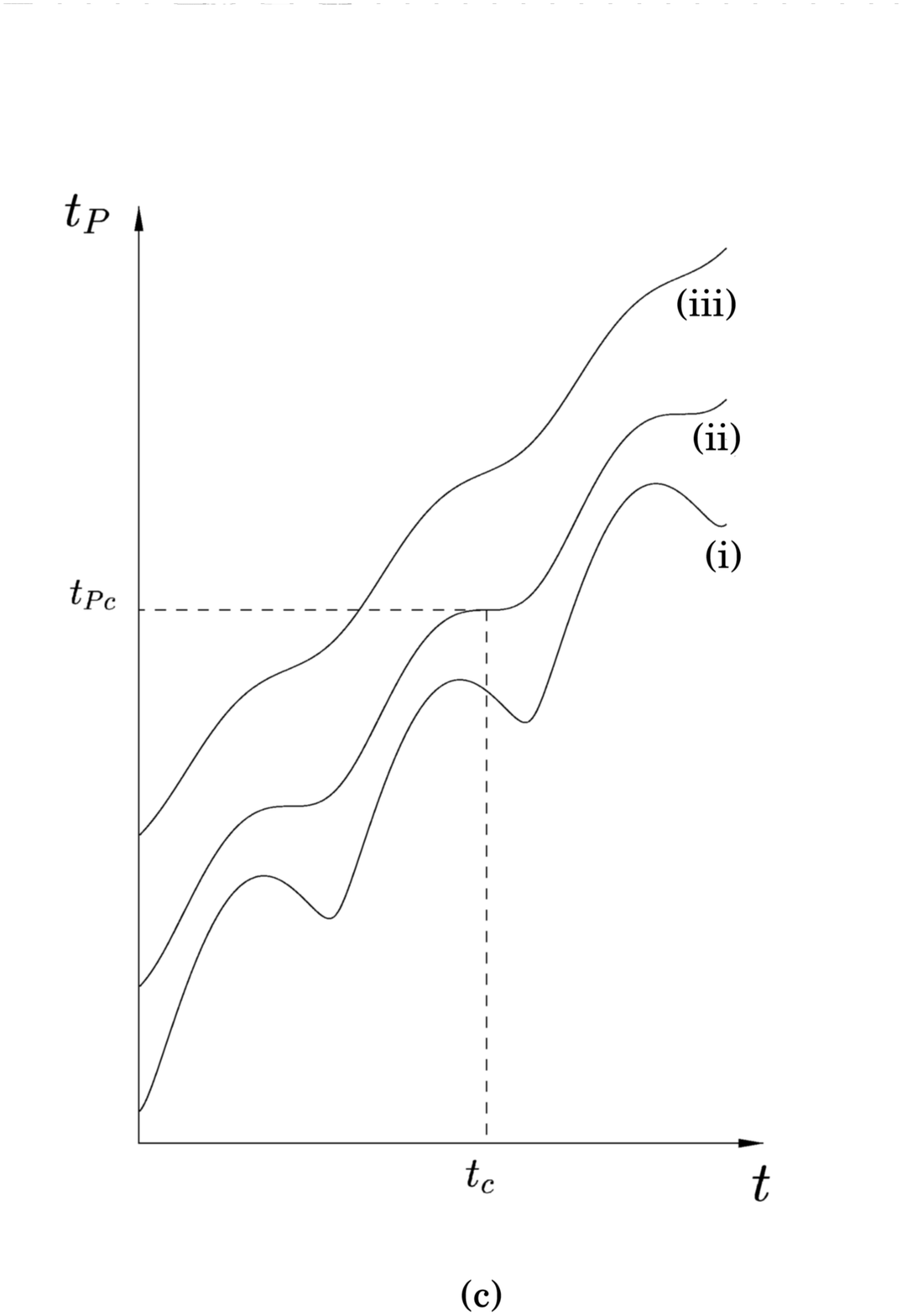}
\caption{(a)~Cross-section
of the \v{C}erenkov-like envelope (bold curves)
of the spherical Huygen's wave fronts (fine circles)
emitted by a small element \point{S} within an
extended, rotating superluminal source
of angular velocity $\omega$.
S is on a circle of radius $r=2.5c/\omega$,
or, in our dimensionless units,
$\hat{r} \equiv r\omega/c = 2.5$;
{\it i.e.} its instantaneous linear velocity is $r\omega=2.5c$.
The cross-section is in the plane of \point{S}'s rotation;
dashed circles designate the light cylinder
$\subP{r} = c/\omega$ ($\subP{\hat r} = 1$)
and the orbit of \point{S}.
(b)~Three-dimensional view of the light cylinder,
the envelope of wave fronts emanating from \point{S},
and the cusp along which the two sheets $\phi_\pm$ of this envelope meet tangentially.
(c)~The relationship
between reception time $\subP{t}$ and
source (retarded) time $t$ [~\Eq{time}] plotted
for $\hat{r}=2.5$ and three different observation points. 
The maxima and minima of curve (i) occur on the
sheets $\phi_\pm$ of the envelope, respectively. Curve (ii)
corresponds to an observation point that is located on the cusp. 
Note that the waves
emitted during an interval of retarded time centered at $t_c$
are received over a much shorter interval of observation time at $t_{P_c}$.
Curve (iii) is for an observation point
that is never crossed by the rotating sheets of the envelope.
(After Ref.~\citeonline{ArdavanH:Speapc}.)}
\label{fig:envelope}
\end{figure}

In Ref.~\citeonline{ArdavanH:Speapc},
the field of a superluminally rotating extended source was evaluated
by superposing the fields of its constituent volume elements,
{\it i.e.}\ by convolving its density with the 
familiar Li\'enard-Wiechert field
of a rotating point source.
This Li\'enard-Wiechert field
is described by an expression essentially identical
to that which is encountered in the analysis of synchrotron radiation,
except that its value at any given observation time
receives contributions from more than one retarded time.
The multivalued nature of the retarded time
is an important feature of all superluminal emission;
we shall begin, therefore,
by describing the relationship between observation (reception) time
and retarded (emission) time
for the particular case of a rotating source
with the aid of \Fig{envelope}.

\subsection{Multivalued retarded times, the cusp and temporal focusing}
Parts (a) and (b) of \Fig{envelope}
show the wave fronts
that emanate from a small, circularly moving superluminal source \point{S}.
As we have already pointed out,
no superluminal source can be truly pointlike.
Here we are considering a volume element
of an extended source
whose linear dimensions are much smaller
than the other length scales of the problem.

The emission of waves
by any moving point source whose speed exceeds the wave speed
is described by a Li\'enard-Wiechert field
that has extended singularities.
These singularities occur on the envelope of wave fronts
where the Huygens wavelets emitted at differing retarded times
interfere constructively and so form caustics.
A well-understood example
is the emission of acoustic waves by a point source
that moves along a straight line with a constant supersonic speed.
In this case, 
 simple caustic forms along a cone issuing from the source,
the so-called Mach cone,
and most of the emitted energy
is confined to the vicinity of this propagating `shock' front.
Another, similar example
is the formation of the \v{C}erenkov cone
in the electromagnetic field
of a uniformly moving point charge
whose speed exceeds the speed of light inside a dielectric medium. 

When the supersonic or superluminal motion of such sources
is in addition accelerated,
the simple conical caustic that occurs in the Mach or \v{C}erenkov radiation
is replaced by a two-sheeted envelope with a cusp.\cite{LilleyGM:Somans,LowsonMV:FochBn,ArdavanH:Genfnd}
The effect of acceleration
is to give rise to a one-dimensional locus of observation points
at which more than two simultaneously received wave fronts meet tangentially.
The spherical wave fronts
that are centered at the retarded positions of the source
neighboring a point from which such coalescing wave fronts emanate
cannot but be mutually tangential (in pairs) to two distinct surfaces,
surfaces that constitute the separate sheets of a cusped envelope.

More specifically,
the \v{C}erenkov-like envelope
that is generated by a uniformly rotating superluminal source
consists of a tube-like surface whose two sheets meet,
and are tangent to one another,
along a spiraling cusp curve;
this envelope is depicted in \Fig{envelope}
and mathematically described in Eqs.~\eqref{varphiP}--\eqref{zeroth} below.
At any given observation time,
three wave fronts pass through an observation point inside the envelope,
while only one wave front passes through a point outside this surface.
The envelope and its cusp are the loci of observation points
at which, respectively,
two or three of the simultaneously received wave 
fronts are tangential to one another.
To specify the retarded times $t$ at which various wave fronts are emitted,
let us adopt a cylindrical coordinate system
based on the axis of rotation and denote the trajectory of the volume element \point{S},
shown in \Fig{envelope},
by
\begin{equation}
r={\rm const}.,\quad\varphi(t) = \hat{\varphi} + \omega t,\quad z={\rm const}.,
\label{eq:varphi}
\end{equation}
where $\hat{\varphi}$ denotes the initial value of $\varphi$,
and $\omega$ is the angular velocity of \point{S}.
Let a stationary observer
be positioned at a point \point{P}, with cylindrical polar coordinates
$(\subP{r}, \subP{\varphi}, \subP{z})$.
The retarded-time separation $R(t)$
between the source volume element and the observer
({\it i.e.} their instantaneous separation
at the time $t$ of emission) will therefore be
\begin{equation}
R(t)=[(\subP{z}-z)^2+\subP{r}^2+r^2
-2r\subP{r}\cos(\subP{\varphi}-\hat{\varphi}-\omega t)]^\frac{1}{2}.
\label{eq:R}
\end{equation}
The relationship between the retarded time $t$
and the observation time $\subP{t}$, {\it i.e.} 
\begin{equation}
\subP{t}=t+\frac{R(t)}{c},
\label{eq:time}
\end{equation}
is plotted in \Fig{envelope}(c)
for the source speed $r\omega=2.5 c$
and for three classes of stationary observation points:
those, located sufficiently close to the plane of rotation,
that are periodically crossed by the two sheets of the rigidly rotating envelope [curve (i)],
or by just the cusp curve of the envelope [curve (ii)],
and those at higher latitudes that are never crossed by the envelope [curve (iii)].

The ordinates of the neighboring extrema of curve (i) in \Fig{envelope}(c)
designate those observation times,
during each rotation period,
at which the two sheets of the envelope go past the stationary observer
[see Eqs.~\eqref{stationary}--\eqref{varphiP} below].  
Thus, the field inside the envelope receives contributions
from three distinct values of the retarded time [curve (i)],
while the field outside the envelope is influenced by only 
a single instant of emission time [curves (i) and (iii)]. 
The constructive interference
of the emitted waves on the envelope
(where two of the contributing retarded times coalesce)
and on its cusp
(where all three of the contributing retarded times coalesce [curve (ii)])
gives rise to the divergence of the Li\'enard-Wiechert field on these loci. 
There is a higher-order focusing of the waves,
and so a higher-order mathematical singularity,
on the cusp than on the envelope itself. 
While the singularity that occurs on the envelope is integrable,
that which occurs on the cusp is not.
In that it occurs in the temporal as well as the spatial domain,
this focusing is distinct
from that produced by a conventional horn, mirror or lens.
The enhanced amplitude on the cusp
is due to the contributions
from emission over an extended period of source time
reaching the observer over a significantly shorter period of observation time.

The Li\'enard-Wiechert field derived in Ref.~\citeonline{ArdavanH:Speapc}
was used as the Green's function
for calculating the emission from a superluminal polarization current,
comprising both poloidal and toroidal components,
whose distribution pattern rotates
(with an angular frequency $\omega$)
and oscillates (with a frequency $\Omega$)
at the same time.\cite{ArdavanH:Speapc}
It was found that the convolution of the density of this current
with the Green's function described above
results in a field that decays nonspherically:
a field whose strength diminishes with the distance 
$\subP{R}$ from the source as $\subP{R}^{-1/2}$,
rather than $\subP{R}^{-1}$,
within the bundle of cusps that emanate from the 
constituent volume elements of the source
and extend into the far zone. 
This result,
which has now been demonstrated 
experimentally,\cite{ArdavanA:Exponr,SingletonJ:Expdes}
was derived in Ref.~\citeonline{ArdavanH:Speapc}
by setting the observation point within the bundle of generated cusps
and evaluating the convolution integrals over various 
dimensions of the source.\cite{ArdavanH:Speapc}
The steps in this procedure are listed below.
\begin{enumerate}
\item
The integration with respect to the azimuthal extent of the source
was performed by means 
of Hadamard's method.\cite{HadamardJ:lecCau,HoskinsRF:GenFun}
It was shown that the Hadamard finite part
of the divergent integral that describes the field
of a superluminally rotating ring with a sinusoidal density distribution
consists of two parts:
one part is exclusively contributed
by the two elements on the ring that approach the observer
along the radiation direction with the speed of light at the retarded time
({\it i.e.} the elements for which $\dif R/\dif t = -c$),
and the other part is contributed by the entire extent of the ring.
\item
The integration
with respect to the radial dimension of the source
was subsequently performed by the method of stationary phase.\cite{BorvikovVA:UnStaPha}
\end{enumerate}
It was found that,
when the radiation frequency is much higher than the rotation frequency $\omega$,
the main contribution toward the field of a superluminally rotating annular ring
comes from the vicinity of the point on the ring
that approaches the observer not only with the wave speed,
but also with zero acceleration
({\it i.e.} the point at which $\dif R/\dif t = -c$
and $\dif^2 R/\dif t^2=0$ simultaneously).

These contributing source elements
are the ones for which the time-domain phase $\subP{t}=t+R(t)/c$ is doubly stationary. 
Differentiating \Eq{time} with respect to $t$,
we can see that
\begin{equation}
\frac{\dif  R}{\dif t} = -c\qquad{\rm and}\qquad\frac{\dif^2 R}{\dif t^2} = 0
\label{eq:Rdot}
\end{equation}
are equivalent to
\begin{equation}
\frac{\dif\subP{t}}{\dif t} = 0\qquad{\rm and}\qquad\frac{\dif^2 \subP{t}}{\dif t^2} = 0.
\label{eq:stationary}
\end{equation}
These conditions jointly define
the point of inflection in curve (ii) of \Fig{envelope}(c),
corresponding to the cusp passing
through the point of observation \point{P}.

\begin{figure}[htbp]
\centering
\includegraphics[width=8.3cm]{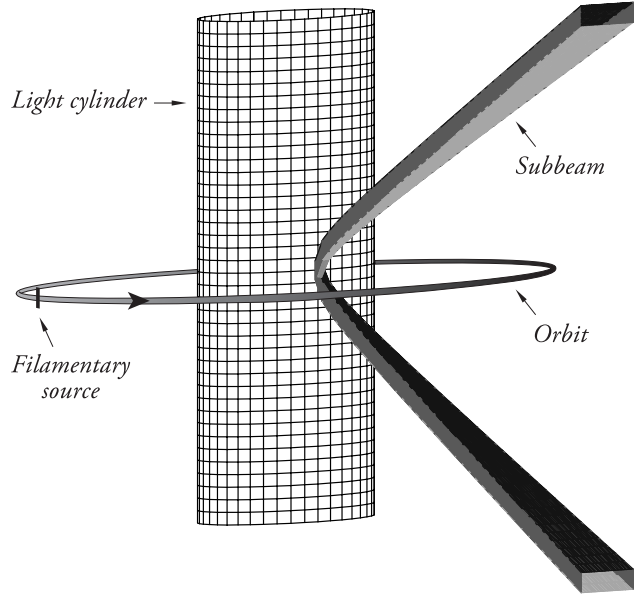}
\caption[The light cylinder and subbeam]%
{Schematic illustration of the light cylinder $r=c/\omega$,
the filamentary part of the source
that approaches the observeration point
with the speed of light and zero acceleration
at the retarded time,
the orbit of this filamentary source,
and the subbeam formed by the bundle of cusps
that emanate from the constituent volume elements of this filament. 
The subbeam is diffractionless in the direction of $\subP{\theta}$.
The figure represents a snapshot
corresponding to a fixed value of the observation time $\subP{t}$. 
The polar width $\delta\subP{\theta}$ of this subbeam
decreases with the distance $\subP{\hat{R}}$
in such a way that the thickness $\subP{\hat{R}}\delta\subP{\theta}$
of the subbeam in the polar direction
remains constant: 
It equals the projection, $\delta\hat{z}\sin\subP{\theta}$,
of the $\hat{z}$ extent, $\delta\hat{z}$, 
of the contributing filamentary source
onto a direction normal to the line of sight. 
The azimuthal width of the subbeam,
on the other hand,
is subject to diffraction as in any other radiation beam:
$\delta\subP{\varphi}$ is independent of $\subP{\hat{R}}$.}
\label{fig:subbeam}
\end{figure}

The collection of volume elements
satisfying \Eq{Rdot} within an extended source
has a filamentary locus
that is approximately parallel to the axis of rotation
for an observation point located in the far zone (\Fig{subbeam}).
The nonspherically decaying field
that is generated by a volume-distributed source
arises almost exclusively from the elements in the vicinity of this narrow filament,
a filament whose position within the source
depends on the location of the observer.

\subsection{The zeroth-order evaluation
of the angular position
of the nonspherically decaying beam\label{sec:zeroth}}
The angle of observation corresponding to the cusp,
and the reason for the filamentary structure
of the contributing parts of the extended source
may be inferred from the above equations. 
Applying the first condition in \Eq{Rdot} to \Eq{R}
and solving the resulting equation for the retarded time $t$,
or equivalently the retarded position $\varphi=\omega t+{\hat\varphi}$,
we obtain
\begin{equation}
\varphi=\varphi_\pm\equiv\subP{\varphi}+2\pi
  -\arccos\bigg(\frac{1\mp\Delta^{1/2}}{\hat{r}\subP{\hat r}}\bigg),
\label{eq:varphipm}
\end{equation}
where
\begin{equation}
\Delta\equiv({\subP{\hat r}}^2-1)(\hat{r}^2-1)-(\hat{z}-\subP{\hat{z}})^2.
\label{eq:Delta}
\end{equation}
In these expressions,
$(\hat{r},\hat{z};\subP{\hat r},\subP{\hat{z}})$
stand for $(r\omega/c, z\omega/c;\subP{r}\omega/c, \subP{z}\omega/c)$,
{\it i.e.} for the coordinates $(r,z;\subP{r},\subP{z})$
of the source point and the observation point
in units of the light-cylinder radius $c/\omega$. 
(This radius, which automatically appears in the present calculations,
turns out to be the main length scale of the problem.)  

The retarded times
$t_\pm\equiv(\varphi_\pm-{\hat\varphi}+2n\pi)/\omega$
respectively represent the maxima and minima of curve (i) in \Fig{envelope}(c)
 where $n$ is an integer.
Applying both conditions of \Eq{Rdot} to \Eq{R},
we obtain \Eq{varphipm} and $\Delta=0$.
The retarded time $t_c\equiv t_\pm\vert_{\Delta=0}$
represents the inflection point of curve (ii) in \Fig{envelope}(c).
Curve (iii) in \Fig{envelope}(c)
corresponds to an observation point for which $\Delta<0$,
and so $\varphi_\pm$ are not real. 

The envelope of wave fronts
comprises those observation points at which two retarded times coalesce,
{\it i.e.} at which $t=t_\pm$.
Inserting these values of the retarded time in \Eq{time}
and solving the resulting equation for $\subP{\varphi}$
as a function of $(\subP{r},\subP{z})$
at a fixed observation time $\subP{t}$,
we find that 
\begin{equation}
\subP{\varphi}=\omega\subP{t}+\hat{\varphi}-\phi_\pm(\subP{r},\subP{z}),
\label{eq:varphiP}
\end{equation}
where
\begin{equation} 
\phi_\pm\equiv\hat{R}_\pm+2\pi
  -\arccos\bigg(\frac{1\mp\Delta^{1/2}}{\hat{r}\subP{\hat r}}\bigg),
\label{eq:phipm}
\end{equation}
with
\begin{equation}
\hat{R}_\pm\equiv
      [(\hat{z}-\subP{\hat{z}})^2+\hat{r}^2+{\subP{\hat r}}^2-2(1\mp\Delta^{1/2})]^{1/2}.
\label{eq:Rhatpm}
\end{equation}
These equations describe a rigidly rotating surface
in the space $(\subP{r},\subP{\varphi},\subP{z})$ of observation points
that extends from the light cylinder $\subP{\hat r}=1$ to infinity
(see \Fig{envelope}).  

The two sheets $\phi_\pm$ of this envelope meet at a cusp. 
The cusp occurs along the curve
\begin{equation}
\Delta=0,\quad
\subP{\varphi}=\omega\subP{t}+\hat{\varphi}-\phi_\pm(\subP{r},
\subP{z})\big\vert_{\Delta=0},  
\label{eq:cuspcurve}
\end{equation}
shown in \Fig{cuspProj}(a).
It can be easily seen that,
for a far-field observation point
with the spherical polar coordinates $\subP{R}\equiv(\subP{r}^2+\subP{z}^2)^{1/2}$,
$\subP{\theta}\equiv\arccos(\subP{z}/\subP{R})$, $\subP{\varphi}$,
\Eq{cuspcurve} reduces to
\begin{equation}
\subP{\theta}=\arcsin(\hat{r}^{-1})+\cdots,\quad
  \subP{\varphi}=\varphi-\smfrac{3}{2}\pi+\cdots,
\label{eq:zeroth}
\end{equation}
to within the zeroth order in the small parameter $\subP{\hat R}^{-1}$,
where $\subP{\hat R}\equiv \subP{R}\omega/c$.
[The higher order terms of this expansion
are given in Eqs.~\eqref{cuspRpinter1Taylor} and \eqref{cuspRpinter2Taylor}.]
In other words,
the cusp that is detected at an observation point
$(\subP{R},\subP{\theta},\subP{\varphi})$
in the far zone
arises from the constructive interference
of the waves that were emitted by the volume elements
at $\hat{r}=\csc\subP{\theta}$, $\varphi=\subP{\varphi}+\smfrac{3}{2}\pi$,
regardless of what their $z$ coordinates may be.
These volume elements therefore have a filamentary locus
parallel to the axis of rotation
whose length is of the order of the $z$ extent of the source distribution
along the line $\hat{r}=\csc\subP{\theta}$,
$\varphi=\subP{\varphi}+\smfrac{3}{2}\pi$
(see \Fig{subbeam}).

\subsection{The filamentary locus of the contributing source elements}
The locus of source elements
that approach the observer with the wave speed and zero acceleration
at the retarded time
has a filamentary shape not only within the zeroth order approximation
in the small parameter $\subP{\hat R}^{-1}$,
but in general.
To demonstrate this,
we need to introduce the notion of {\it bifurcation surface}.%
\cite{ArdavanH:Genfnd}.

\begin{figure}[htbp]
\centering
\includegraphics[angle=-90,width=0.6\linewidth]{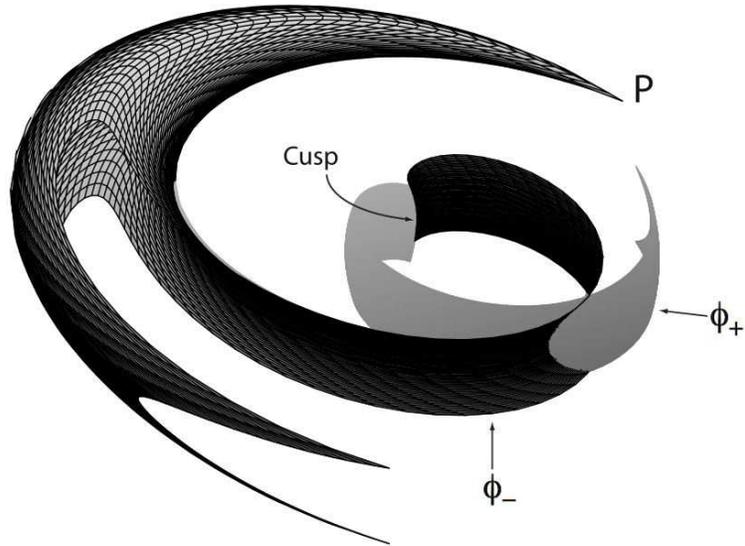}
\caption[The bifurcation surface]%
{The bifurcation surface of the observation point P
for a source whose rotational motion is counterclockwise.
The source points that lie inside this surface
influence the field at P
at three distinct values of the retarded time,
while those that lie outside this surface
influence the field at only a single value of the retarded time.
The source elements
on the filamentary locus at which the cusp curve of this surface
intersects the source distribution approach P
with the speed of light and zero acceleration at the retarded time
and so generate a nonspherically decaying field at P.}
\label{fig:bifurcation}
\end{figure}

When deriving the equation describing the envelope of wave fronts,
we kept the coordinates $(r,{\hat\varphi},z)$,
which label a rotating source element,
fixed
and found the surface in the space $(\subP{r},\subP{\varphi},\subP{z})$
of observation points on which $\dif R/\dif t=-c$
at a given time $\subP{t}$.
If we keep  $(\subP{r},\subP{\varphi},\subP{z})$ and $\subP{t}$ fixed,
then $\dif R/\dif t=-c$ would describe a surface
that resides in the space $(r,\varphi,z)$ of source points:
the so-called bifurcation surface of the observation point \point{P}.
Like the envelope,
the bifurcation surface consists of two sheets
that meet tangentially along a cusp
(a spiraling curve on which $\dif^2 R/\dif t^2=0$),
but the bifurcation surface issues from the observation point \point{P}
(rather than the source point \point{S})
and spirals about the rotation axis in the opposite direction
to the envelope (see \Fig{bifurcation}).
The similarity between the two surfaces
stems from the following reciprocity properties of \point{P} and \point{S}:
the equation describing the envelope, \Eq{varphiP},
remains invariant under the interchanges
$r\leftrightarrow\subP{r},z\leftrightarrow\subP{z},
\varphi\leftrightarrow -\subP{\varphi},t\leftrightarrow -\subP{t}$.

The locus of source elements
that approach an observer \point{P} with the wave speed and zero acceleration
at the retarded time
is given by the intersection of the cusp curve of the bifurcation surface of \point{P}
with the volume of the source.
This filamentary locus
has exactly the same shape as the cusp curve of the envelope
[shown in \Fig{cuspProj}(a)],
except that it resides in the space of source points,
instead of the space of observation points,
and points in the direction of the source velocity. 
The projection of this curve onto the $(r, z)$ plane
consists of a branch of a hyperbola
with asymptotes that lie along the angles $\arcsin(\subP{\hat{r}}^{-1})$
and $\pi-\arcsin(\subP{\hat{r}}^{-1})$
with respect to the $z$ axis
[see \Fig{cuspProj}(b)].
For an observation point that is located in the far zone, therefore,
the projection of the cusp curve of the bifurcation surface
onto the $(r, z)$ plane
is virtually parallel to the rotation axis. 

The reciprocity relations referred to above
ensure that if a source element \point{S}
is located on the cusp curve of the bifurcation surface of an observer \point{P},
then the envelope of the wave fronts emitted by \point{S}
would have a cusp passing through \point{P}
(or, conversely,
if an observer \point{P} is located on the cusp curve of the envelope
of wave fronts emitted by a source element \point{S},
then the cusp curve of the bifurcation surface of \point{P} would pass through \point{S}).
In the case of a single point source,
the retarded position $\varphi$ of the source linearly changes with time
($\varphi={\hat\varphi}+\omega t$),
and so the cusp that it generates is both spiral-shaped
and rigidly rotates about the $z$ axis.
In the case of an extended source, on the other hand,
the position $\varphi$ of each contributing source element
(an element that lies on the cusp curve of the bifurcation surface of a far-field observer \point{P})
is fixed ($\varphi=\subP{\varphi}+3\pi/2$,
$\hat{r}=\csc\subP{\theta}$),
and the elements that occupy that position are constantly changing.
The cusps generated by the moving source elements
that pass through this fixed position at various retarded times
have a locus, at any given observation time,
that is straight and stationary as shown in \Fig{subbeam}.
In other words, the source elements
constituting the filament at $\varphi=\subP{\varphi}+3\pi/2$,
$\hat{r}=\csc\subP{\theta}$,
each contribute a quasi-instantaneous `pulse'
of nonspherically decaying electromagnetic radiation
that in the far field appears to have propagated out 
along a virtually straight-line locus
defined by the angle $\subP{\theta}=\arcsin({\hat r}^{-1})$.

\subsection{Objectives and organization of the paper}
The objectives of the present paper are as follows
(the location of the resolution of each objective is given in
brackets):
\begin{enumerate}
\item
to show that the nonspherical decay of the radiation field
that arises from a rotating superluminal source
remains in force at all distances from this source
independently of the frequency of the radiation (\Section{frequency});
\item
to specify the dimensions
of the filamentary part of the source
that makes the main contribution
toward the value of the nonspherically decaying field
[Eqs.~\eqref{deltar}--\eqref{deltavarphi}];
\item
to show that the bundle of cusps
emanating from this filament
delineates a radiation `subbeam'
that is nondiffracting in one dimension;
that is to say,
the width of this beam in the polar direction
remains the same at all distances from the source 
[\Fig{subbeam}, \Eq{deltathetapApprox}];
\item
to clarify how the requirements of conservation of energy
are met by the nonspherically decaying radiation: 
the cross-sectional area of each nondiffracting subbeam
increases as $\subP{R}$,
rather than $\subP{R}^2$,
with the distance $\subP{R}$ from the source (\Section{ndsubbeams});
and
\item
to show that the overall radiation beam
within which the field decays nonspherically
consists, in general,
of an incoherent superposition of the coherent nondiffracting 
subbeams described above (\Section{ndsubbeams}).
\end{enumerate}

We begin with the mathematical formulation of the problem in \Section{nsdecay}.
In \Section{asymptotic}, we show that objectives 1 and 2 can be achieved
by replacing the method of stationary phase
used in Ref.~\citeonline{ArdavanH:Speapc}
with the method of steepest descents.~\cite{BleisteinN:Asei}
By converting the Fourier-type integral
over the radial extent of the source
to a Laplace-type integral
and making use of contour integration,
we present an asymptotic analysis
for which the large parameter is the distance from the source
(in units of the light-cylinder radius $c/\omega$)
rather than the radiation frequency. 
Not only is there no restriction
on the range of frequencies
for which the emission from a rotating superluminal source decays nonspherically, 
but the more distant the observation point,
the more accurate the asymptotic analysis that predicts this decay rate.

The more poweful asymptotic technique
we employ here establishes, moreover,
that the transverse dimensions of the filamentary part of the source
responsible for the nonspherically decaying field
are of the order of $\delta\hat{r}\propto{\subP{\hat{R}}}^{-2}$ in the radial direction
and $\delta\varphi\propto{\subP{\hat{R}}}^{-3}$ in the azimuthal direction
(see \Section{ndsubbeams}).
The dimension of this filament in the direction parallel to the rotation axis
is of the order of the length scale of the source distribution in that direction.

The corresponding dimensions of the bundle of cusps
that emanate from the contributing source elements
can be easily inferred
from the above dimensions of the filamentary region containing these elements.
The cusps occupy a solid angle
in the space of observation points
whose azimuthal width $\delta\subP{\varphi}$ has a constant value
(as does a conventional radiation beam)
but whose polar width $\delta\subP{\theta}$
decreases with the distance $\subP{R}$ as $\subP{R}^{-1}$.
This may be seen by considering a
cohort of propagating polarization-current volume elements
that are at the same azimuthal angle $\varphi$ and radius $r$
(possessing the same speed $r\omega$)
but at differing heights $z$.
Each will give rise to a cusp in the far zone
that forms the angle $\subP{\theta}=\arcsin({\hat r}^{-1})$ with the $z$ axis,
but starts from a different height at the light cylinder
(see \Fig{subbeam}). 
The spatial extent in the direction
of increasing $\subP{\theta}$
of the composite set of cusps
from this cohort of volume elements (the subbeam)
will therefore be determined solely by the height $\delta z$
of the region confining the polarization current.
Projected onto a direction normal to the line of sight,
this will result in a width $w=\vert\delta z\vert \sin \subP{\theta}$ 
occupied by the cusps
that is independent of the distance $R_P$ from the source.
(Note that $w$ is a fixed {\em linear} width,
rather than an {\em angular} width.)

Thus, the area $\subP{R}^2\sin\subP{\theta}\delta\subP{\theta}\subP{\varphi}$
subtended by the bundle of cusps defining this subbeam
increases as $\subP{R}$, rather than $\subP{R}^2$,
with the distance $\subP{R}$ from the source.
In order that the flux of energy
remain the same across a cross section of the subbeam, therefore,
it is essential that the Poynting vector associated with this radiation
correspondingly decay as $\subP{R}^{-1}$,
rather than $\subP{R}^{-2}$. 
This requirement is, of course, met automatically
by the radiation that propagates along the nondiffracting subbeam.

For a rotating superluminal source
with the radial boundaries $\hat{r}_<>1$ and $\hat{r}_>>\hat{r}_<$,
the nonspherically decaying radiation is detectable in the far zone
only within the conical shell
\begin{equation}
\arcsin(1/\hat{r}_>)\le\subP{\theta}\le\arcsin(1/\hat{r}_<).
\label{eq:conicalShell}
\end{equation}
These limits on $\subP{\theta}$
merely reflect the fact that a rigidly rotating
extended source with finite radial spread
entails a limited range of linear speeds $r\omega$;
\Eq{zeroth} shows that a limited range of speeds
results in a limited spread in the angular positions of the generated subbeams.
The overall beam described by \Eq{conicalShell} consists, in general,
of a superposition of nondiffracting subbeams
with widely differing amplitudes and phases.
The individual subbeams
(which would be narrower and more distinguishable,
the further away is the observer from the source)
decay nonspherically,
but the incoherence of their phase relationships
ensures that the integrated flux of energy
associated with their superposition across this finite solid angle
remains independent of $\subP{R}$.   

Having made a preliminary description
of the salient features of the analysis,
we now embark on the detailed treatment of the problem
in Sections \ref{sec:nsdecay} to \ref{sec:ndsubbeams}.
We conclude in \Section{concluding}
with some remarks on the applicability of our analysis
to numerical calculations of the emission from superluminal sources 
and to the observational data on the giant pulses received from pulsars.

\section{The nonspherically decaying component of the radiation field
from a rotating superluminal source\label{sec:nsdecay}}

As in Ref.~\citeonline{ArdavanH:Speapc},
we base our analysis
on a polarization current density ${\bf j}=\partial{\bf P}/\partial t$
for which
\begin{equation}
P_{r,\varphi,z}(r,\varphi,z,t)
= s_{r,\varphi,z}(r,z)\cos(m\hat{\varphi})\cos(\Omega t),
\qquad -\pi<\hat{\varphi}\le\pi,
\label{eq:elecPol}
\end{equation}
with
\begin{equation}
\hat{\varphi}\equiv\varphi-\omega t,
\label{eq:varphiHat}
\end{equation}
where $P_{r,\varphi,z}$ are the components of the polarization ${\bf P}$
in a cylindrical coordinate system based on the axis of rotation,
${\bf s}(r,z)$ is an arbitrary vector
that vanishes outside a finite region of the $(r,z)$ space,
and $m$ is a positive integer.
For a fixed value of $t$,
the azimuthal dependence of the density \eqref{elecPol}
along each circle of radius $r$ within the source
is the same as that of a sinusoidal wave train with the wavelength $2\pi r/m$
whose $m$ cycles fit around the circumference of the circle smoothly.
As time elapses,
this wave train both propagates around each circle with the velocity $r\omega$
and oscillates in its amplitude with the frequency $\Omega$.
This is a generic source: 
One can construct any distribution with a uniformly rotating pattern,
$P_{r,\varphi,z}(r,\hat{\varphi},z)$,
by the superposition over $m$
of terms of the form $s_{r,\varphi,z}(r,z,m)\cos(m\hat{\varphi})$.

The electromagnetic fields
\begin{equation}
{\bf E}=-\subP{{\bf\nabla}} A^0 - \frac{\partial{\bf A}}{\partial(c \subP{t})},
\quad {\bf B}=\subP{{\bf\nabla}}{\bf\times A},
\label{eq:fields}
\end{equation}
that arise from such a source are given,
in the absence of boundaries,
by the following classical expression for the retarded four-potential:
\begin{equation}
A^\mu(\subP{{\bf x}},\subP{t})=c^{-1}\int\dif^3 x\dif t\, j^\mu({\bf x},t)\delta(\subP{t}-t-R/c)/R
\quad \mu=0,\cdots,3
\label{eq:Amu}
\end{equation}
Here,
$(\subP{{\bf x}},\subP{t})=(\subP{r},\subP{\varphi},\subP{z},\subP{t})$ and $({\bf x},t)=(r,\varphi,z,t)$
are the space-time coordinates of the observation point and the source points, respectively,
$R$ stands for the magnitude of ${\bf R}\equiv\subP{{\bf x}}-{\bf x}$,
and $\mu=1,2,3$ designate the spatial components,
${\bf A}$ and ${\bf j}$,
of $A^\mu$ and $j^\mu$ in a Cartesian coordinate system.\cite{JacksonJD:Classical}

In Ref.~\citeonline{ArdavanH:Speapc},
we first calculated the Li\'enard-Wiechert field
that arises from a circularly moving point source
(representing a volume element of an extended source)
with a superluminal speed $r\omega>c$,
{\it i.e.}\ considered a generalization
of the synchrotron radiation to the superluminal regime.
We then evaluated the integral representing the retarded field
(rather than the retarded potential)
of the extended source \eqref{elecPol}
by superposing the fields
generated by the constituent volume elements of this source,
{\it i.e.}\ by using the generalization of the synchrotron field
as the Green's function for the problem
(see also Ref.~\citeonline{ArdavanH:Speapc1}).
In the superluminal regime,
this Green's function has extended singularities,
singularities that arise from the constructive intereference
of the emitted waves on the envelope of wave fronts and its cusp.

Labeling each element of the extended source \eqref{elecPol}
by its Lagrangian coordinate $\hat{\varphi}$
and performing the integration with respect to $t$ and $\hat{\varphi}$
(or equivalently $\varphi$ and $\hat{\varphi}$)
in the multiple integral implied by Eqs.~\eqref{elecPol}--\eqref{Amu},
we showed in Ref.~\citeonline{ArdavanH:Speapc}
that the resulting expression for the radiation field ${\bf B}$ (or ${\bf E}$)
consists of two parts: a part whose magnitude decays spherically,
as $\subP{R}^{-1}$, with the distance $\subP{R}$ from the source
(as in any other conventional radiation field),
and another part ${\bf B}^{\rm ns}$,
with ${\bf E}^{\rm ns}={\hat{\bf n}}{\bf\times}{\bf B}^{\rm ns}$
whose magnitude decays as $\subP{R}^{-1/2}$
within the conical shell described by \Eq{conicalShell}.
(Here, ${\hat{\bf n}}\equiv{\bf R}/R$ is a unit vector in the radiation direction.)

\begin{figure}[htbp]
\centering
\includegraphics[width=8.3cm]{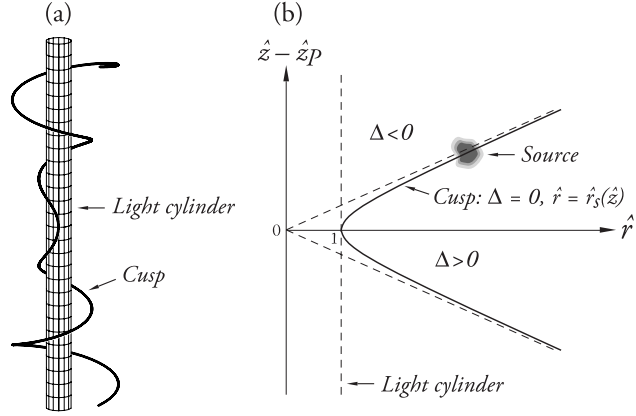}
\caption[A segment of the cusp]%
{(a) A segment of the cusp
of the envelope of wave fronts
emitted by a rotating point source
with the speed $r\omega=3c$. 
This curve is tangent to the light cylinder
at the point
($\subP{\hat r}=1$, $\subP{\varphi}=\varphi-3\pi/2$, $\subP{\hat{z}}=\hat{z}$)
on the plane of the orbit
and spirals outward into the far zone.
Note that this figure represents a snapshot
at a fixed value of the observation time $\subP{t}$. 
The cusp curve of the bifurcation surface of an observer \point{P}
shown in \Fig{bifurcation}
has precisely the same shape,
except that it resides in the space of source points,
instead of the space of observation points,
and spirals in the counterclockwise direction:
It is tangent to the light cylinder
at the point ($\hat{r}=1$, $\varphi=\subP{\varphi}+3\pi/2$, $\hat{z}=\subP{\hat{z}}$).
(b) The projections of the cusp curve of the bifurcation surface
and a localized source distribution
onto the $(\hat{r}, \hat{z})$ plane.
Only the part of the source
that lies close to the cusp in $\Delta>0$
contributes to the nonspherically decaying radiation.
The source elements whose $(\hat{r},\hat{z})$ coordinates fall in $\Delta<0$
approach the observer with a speed $\dif R/\dif t<c$
at the retarded time
and so make contributions toward the field
that are no different from those made in the subluminal regime. 
The asymptotes of the hyperbola $\Delta=0$
make the angles $\arcsin(1/\subP{\hat r})$ and $\pi-\arcsin(1/\subP{\hat r})$
with the $z$ axis,
so that for an observation point in the far zone ($\subP{\hat r}\gg1$)
the projection of the cusp onto the $(\hat{r}, \hat{z})$ plane
is (as depicted in \Fig{subbeam})
effectively parallel to the rotation axis.}
\label{fig:cuspProj}
\end{figure}

The expression found in Ref.~\citeonline{ArdavanH:Speapc} [Eq.~(47)]
for the nonspherically decaying component
of the field within this conical shell, in the far zone,
is
\begin{eqnarray}
{\bf B}^{\rm ns}\simeq & -\smfrac{4}{3}\im\exp[\im(\Omega/\omega)(\subP{\varphi}+3\pi/2)]
    \sum_{\mu=\mu_\pm}\mu\exp(-\im\mu\subP{\hat{\varphi}}) \nonumber \\
  & \times\sum_{j=1}^3 \bar{q}_j
     \int_{\Delta\ge 0}\hat{r}\dif\hat{r}\,\dif\hat{z}
     \,\Delta^{-1/2}{\bf u}_j\exp(-\im\mu\phi_-),
\label{eq:Bns} \\
\end{eqnarray}
where
\begin{equation}
\mu_\pm\equiv(\Omega/\omega)\pm m,
\label{eq:mu}
\end{equation}
\begin{equation}
\subP{\hat{\varphi}}\equiv\subP{\varphi}-\omega \subP{t},
\label{eq:phihatP}
\end{equation}
\begin{equation}
\bar{q}_j\equiv(1\;\quad-\im\Omega/\omega\quad\im\Omega/\omega),
\label{eq:qbar}
\end{equation}
and
\begin{eqnarray}
    & {\bf u}_j \equiv \left\{
      \begin{array}{c}
      s_r\cos\subP{\theta}{\hat{\bf e}}_\parallel+s_\varphi {\hat{\bf e}}_\perp \\
      -s_\varphi\cos\subP{\theta}{\hat{\bf e}}_\parallel+s_r{\hat{\bf e}}_\perp \\
      -s_z\sin\subP{\theta}{\hat{\bf e}}_\parallel
\end{array} \right\},
\label{eq:u_j}
\end{eqnarray}
with $j=1,2,3$.  In the above expression, ${\hat{\bf e}}_\parallel\equiv{\hat{\bf e}}_{\subP{\varphi}}$
(which is parallel to the plane of rotation)
and ${\hat{\bf e}}_\perp\equiv{\hat{\bf n}}{\bf\times}{\hat{\bf e}}_\parallel$
comprise a pair of unit vectors
normal to the radiation direction ${\hat{\bf n}}$.
The domain of integration in \Eq{Bns}
consists of the part of the source distribution ${\bf s}(r,z)$
that falls within $\Delta\ge0$ (see \Fig{cuspProj}).

The function $\phi_-(\hat{r},\hat{z})$
that appears in the phase of the integrand in \Eq{Bns}
is stationary as a function of $\hat{r}$ at
\begin{equation}
\hat{r}=\subP[C]{\hat{r}}(\hat{z})\equiv\{
  \smfrac{1}{2}({\subP{\hat r}}^2+1)-[\smfrac{1}{4}({\subP{\hat r}}^2-1)^2
    -(\hat{z}-\subP{\hat{z}})^2]^{1/2}\}^{1/2}.
\label{eq:rHatC}
\end{equation}
When the observer is located in the far zone,
this isolated stationary point coincides with the locus,
\begin{equation}
\hat{r}=\subP[S]{\hat{r}}\equiv[1+(\hat{z}-\subP{\hat{z}})^2/({\subP{\hat r}}^2-1)]^{1/2},
\label{eq:rHatS}
\end{equation}
of source points that approach the observer with the speed of light and zero acceleration
at the retarded time,
{\it i.e.}\ with the projection of the cusp curve of the bifurcation surface
onto the $(r,z)$ plane (see \Fig{cuspProj}).
For $\subP{\hat{R}}\gg1$,
the separation $\subP[C]{\hat{r}}-\subP[S]{\hat{r}}$ vanishes as ${\subP{\hat{R}}}^{-2}$
[see \Eq{limxiS} below]
and both $\subP[C]{\hat{r}}$ and $\subP[S]{\hat{r}}$ assume the value $\csc\subP{\theta}$.

It follows from \Eq{phipm} that
\begin{equation}
\phi_-\vert_{\hat{r}=\subP[C]{\hat{r}}}\equiv\subP[C]{\phi}=
  \subP[C]{\hat{R}}+\subP[C]{\varphi}-\subP{\varphi},
\label{eq:phi_minusOnrHatC}
\end{equation}
$\partial\phi_-/\partial\hat{r}\vert_{\hat{r}=\subP[C]{\hat{r}}}=0$,
and
\begin{equation}
\left.\frac{\partial^2\phi_-}{\partial\hat{r}^2}\right|_{\hat{r}=\subP[C]{\hat{r}}}\equiv
  a=-{\subP[C]{\hat{R}}}^{-1}[({\subP{\hat r}}^2-1)({\subP[C]{\hat{r}}}^2-1)^{-1}-2],
\label{eq:d2phimdr2onrHatC}
\end{equation}
where
\begin{equation}
\subP[C]{\varphi}=\subP{\varphi}+2\pi-\arccos(\subP[C]{\hat{r}}/\subP{\hat r})
\label{eq:varphiC}
\end{equation}
and
\begin{equation}
\subP[C]{\hat{R}}=\subP[C]{\hat{r}}({\subP{\hat r}}^2-{\subP[C]{\hat{r}}}^2)^{1/2}.
\label{eq:RhatC}
\end{equation}
Note that for observation points of interest to us
(the observation points located outside the plane of rotation, $\subP{\theta}\neq\pi/2$,
in the far zone, $\subP{\hat{R}}\gg1$),
the parameter $a$ has a value whose magnitude increases with increasing $\subP{\hat{R}}$: 
\begin{equation}
a\simeq-\subP{\hat{R}}\sin^4\subP{\theta}\sec^2\subP{\theta}
\label{eq:aparam}
\end{equation}
[see \Eq{d2phimdr2onrHatC}].
In other words,
the phase function $\phi_-$ is more peaked at its maximum,
the farther the observation point is from the source.

This property of the phase function $\phi_-$
distinguishes the asymptotic analysis
that will be presented in the following section
from those commonly encountered in radiation theory.
What turns out to play the role of a large parameter in this asymptotic expansion
is distance ($\subP{\hat{R}}$), not frequency ($\mu_\pm\omega$).

\section{Asymptotic analysis of the integral representing the field for large distance%
\label{sec:asymptotic}}

\subsection{Transformation of the phase of the integrand into a canonical form}

The first step in the asymptotic analysis
of the integral that appears in \Eq{Bns}
is to introduce a change of variable $\xi=\xi(\hat{r},\hat{z})$
that replaces the original phase $\phi_-$ of the integrand
by as simple a polynomial as possible.
This transformation should be one-to-one
and should preserve the number and nature of the stationary points of the phase.%
\cite{BorvikovVA:UnStaPha,BleisteinN:Asei}
Since $\phi_-$ has a single isolated stationary point
at $\hat{r}=\subP[C]{\hat{r}}(\hat{z})$,
it can be cast into a canonical form
by means of the following transformation:
\begin{equation}
\phi_-(\hat{r}, \hat{z})=\subP[C]{\phi}(\hat{z})+\smfrac{1}{2}a(\hat{z})\xi^2,
\label{eq:phi_minus}
\end{equation}
in which $a$ is the coefficient given in Eqs.~\eqref{d2phimdr2onrHatC} and \eqref{aparam}.

The integral in \Eq{Bns} can thus be written as
\begin{equation}
\int_{\Delta\ge0}\hat{r}\dif\hat{r}\,\dif\hat{z}
  \,\Delta^{-1/2}{\bf u}_j\exp(-\im\mu\phi_-)
=\int_{\xi\ge\subP[S]{\xi}}\dif\hat{z}\dif\xi\, A(\xi,\hat{z})\exp(\im\alpha\xi^2),
\label{eq:intDeltaNonneg}
\end{equation}
in which 
\begin{equation}
A(\xi,\hat{z})\equiv
  \hat{r}\Delta^{-1/2}{\bf u}_j\frac{\partial\hat{r}}{\partial\xi}\exp(-\im\mu\subP[C]{\phi}),
\label{eq:AxizHat}
\end{equation}
with
\begin{equation}
\frac{\partial\hat{r}}{\partial\xi}=a\xi\hat{r}\hat{R}_-(\hat{r}^2-1-\Delta^{1/2})^{-1},
\label{eq:drHatdxi}
\end{equation}
and $\alpha\equiv-\mu a/2$.
The stationary point $\hat{r}=\subP[C]{\hat{r}}$
and the boundary point $\hat{r}=\subP[S]{\hat{r}}$
respectively map onto $\xi=0$ and
\begin{equation}
\xi=\subP[S]{\xi}\equiv-[2a^{-1}(\subP[S]{\phi}-\subP[C]{\phi})]^{1/2}
\label{eq:xiS}
\end{equation}
where
\begin{equation}
\subP[S]{\phi}\equiv\phi_-\vert_{\hat{r}=\subP[S]{\hat{r}}}
=2\pi-\arccos[1/(\subP[S]{\hat{r}}\subP{\hat r})]+({\subP[S]{\hat{r}}}^2{\subP{\hat r}}^2-1)^{1/2}.
\label{eq:phiS}
\end{equation}
The upper limit of integration in \Eq{intDeltaNonneg}
is determined by the image of the support
of the source density (${\bf s}$ in ${\bf u}_j$)
under the transformation (\ref{eq:phi_minus}).

The Jacobian $\partial\hat{r}/\partial\xi$
of the above transformation
is indeterminate at $\xi=0$.
Its value at this critical point has to be found
by repeated differentiation of \Eq{phi_minus} with respect to $\xi$,
\begin{equation}
\frac{\partial\phi_-}{\partial\hat{r}}\frac{\partial\hat{r}}{\partial\xi}=a\xi,
\label{eq:repDif1}
\end{equation}
\begin{equation}
\frac{\partial^2\phi_-}{\partial\hat{r}^2}\left(\frac{\partial\hat{r}}{\partial\xi}\right)^2
  +\frac{\partial\phi_-}{\partial\hat{r}}\frac{\partial^2\hat{r}}{\partial\xi^2}=a,
\label{eq:repDif2}
\end{equation}
and the evaluation of the resulting relation \eqref{repDif2}
at $\hat{r}=\subP[C]{\hat{r}}$
with the aid of \Eq{d2phimdr2onrHatC}.
This procedure,
which amounts to applying l'H\^opital's rule,
yields $\partial\hat{r}/\partial\xi\vert_{\xi=0}=1$:
a result that could have been anticipated
in light of the coincidence of transformation (\ref{eq:phi_minus})
with the Taylor expansion of $\phi_-$ about $\hat{r}=\subP[C]{\hat{r}}$
to within the leading order.
Correspondingly,
the amplitude $A(\xi)$ that appears in \Eq{AxizHat}
has the value
$\subP[C]{\hat{r}}({\subP[C]{\hat{r}}}^2-1)^{-1}{\bf u}_j\vert_{\hat{r}=
\subP[C]{\hat{r}}}\exp(-\im\mu\subP[C]{\phi})$
at the critical point \point{C}.

When the observer is located in the far field ($\subP{\hat{R}}\gg1$),
the phase of the integrand on the right-hand side of \Eq{intDeltaNonneg}
is rapidly oscillating
irrespective of how low the harmonic numbers $\mu_\pm$
({\it i.e.}\ the radiation frequencies $\mu_\pm\omega$)
may be.
The leading contribution
to the asymptotic value of integral \eqref{intDeltaNonneg}
from the stationary point $\xi=0$
can therefore be determined by the method of stationary phase.\cite{BorvikovVA:UnStaPha}
However, in the limit $\subP{\hat{R}}\to\infty$, $\subP[S]{\xi}$ reduces to
\begin{equation}
\subP[S]{\xi}\simeq-3^{-1/2}\cos^4\subP{\theta}\csc^5\subP{\theta}{\subP{\hat{R}}}^{-2},
\label{eq:limxiS}
\end{equation}
so that the stationary point $\xi=0$
is separated from the boundary point $\xi=\subP[S]{\xi}$
by an interval of the order of ${\subP{\hat{R}}}^{-2}$ only.
To determine the extent of the interval in $\hat{r}$
from which the dominant contribution toward the value of the radiation field arises,
we therefore need to employ a more powerful technique
for the asymptotic analysis of integral \eqref{intDeltaNonneg},
a technique that is capable of handling the contributions
from both $\subP[C]{\hat{r}}$ and $\subP[S]{\hat{r}}$ simultaneously.

\subsection{Contours of steepest descent\label{sec:contours}}

The technique we shall employ for this purpose
is the method of steepest descents.\cite{BleisteinN:Asei}
We regard the variable of integration in 
\begin{equation}
I(\hat{z})\equiv\int_{\subP[S]{\xi}}^{\xi_>}\dif\xi\, A(\xi,\hat{z})\exp(\im\alpha\xi^2)
\label{eq:IofzHat}
\end{equation}
as complex,
{\it i.e.}\ write $\xi=u+\im v$,
and invoke Cauchy's integral theorem
to deform the original path of integration
into the contours of steepest descent
that pass through each of the critical points
$\xi=\subP[S]{\xi}$, $\xi=0$, and $\xi=\xi_>$.
Here, we have introduced the real variable $\xi_>(\hat{z})$
to designate the image of $\hat{r}_>$ under transformation (\ref{eq:phi_minus}),
{\it i.e.}\ the boundary of the support of the source term ${\bf u}_j$
that appears in the amplitude $A(\xi,\hat{z})$.
We shall only treat the case in which $\mu$ (and hence $\alpha$) is positive;
$I(\hat{z})$ for negative $\mu$ can then be obtained
by taking the complex conjugate of the derived expression
and replacing $\subP[C]{\phi}$ with $-\subP[C]{\phi}$ [see \Eq{AxizHat}].

\begin{figure}[htbp]
\centering
\includegraphics[width=8.3cm]{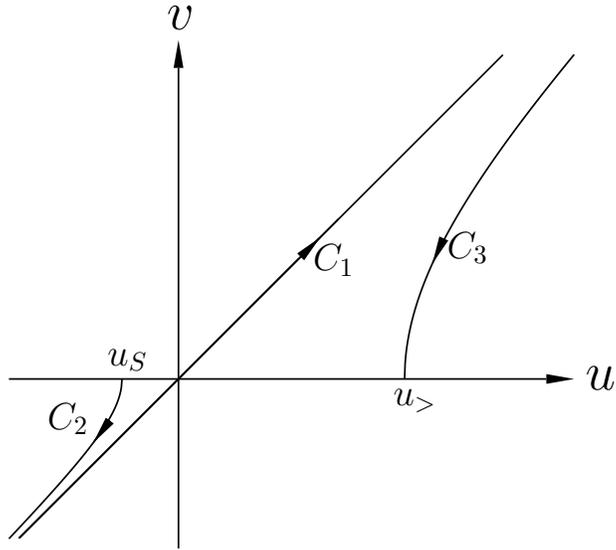}
\caption[The integration contours in the complex plane $\xi=u+\im v$]%
{The integration contours in the complex plane $\xi=u+\im v$.
The critical point \point{C} lies at the origin,
and $\subP[S]{u}$ and $u_>$ are the images under transformation (\ref{eq:phi_minus})
of the radial boundaries $\hat{r}=\subP[S]{\hat{r}}(\hat{z})$
and $\hat{r}=\hat{r}_>(\hat{z})$
of the part of the source that lies within $\Delta>0$
(see \Fig{cuspProj}).
The contours $C_1$, $C_2$, and $C_3$
are the paths of steepest descent
through the stationary point \point{C}
and the lower and upper boundaries of the integration domain,
respectively.}
\label{fig:intCont}
\end{figure}

The path of steepest descent
through the stationary point \point{C}
at which $\xi=0$ is given,
according to
\begin{equation}
\im\xi^2=-2uv+\im(u^2-v^2),
\label{eq:ixi2}
\end{equation}
by $u=v$ when $\alpha$ is positive.
If we designate this path by $C_1$
(see \Fig{intCont}),
then
\begin{eqnarray}
\lefteqn{\int_{C_1}\dif\xi A(\xi,\hat{z})\exp(\im\alpha\xi^2)=
    (1+\im )\int_{-\infty}^{\infty}\dif v A\vert_{\xi=(1+\im )v}\exp(-2\alpha v^2)}
    \nonumber \\
  & & \simeq(2\pi/\mu)^{1/2}\exp[-\im(\mu\subP[C]{\phi}-\pi/4)]{\bf u}_j\subP[C]{\vert}
      \csc\subP{\theta}\vert\sec\subP{\theta}\vert{\subP{\hat{R}}}^{-1/2},
\label{eq:intC1}
\end{eqnarray}
for $\subP{\hat{R}}\gg1$.
Here, we have obtained the leading term
in the asymptotic expansion of the above integral
for large $\subP{\hat{R}}$
by approximating $A\vert_{\xi=(1+\im )v}$
by its value at \point{C}, where $v=0$,
and using \Eq{aparam} to replace $\alpha$
by its value in the far zone.
Note that the next term in this asymptotic expansion
is by a factor of order ${\subP{\hat{R}}}^{-1/2}$
smaller than this leading term.

The path of steepest descent through the boundary point \point{S},
at which $u\equiv \subP[S]{u}=\subP[S]{\xi}$ and $v=0$
[see Eqs.~\eqref{xiS} and \eqref{limxiS}],
is given by $u=-(v^2+{\subP[S]{u}}^2)^{1/2}$,
{\it i.e.}\ by the contour designated as $C_2$ in \Fig{intCont}.
The real part of 
\begin{equation}
\im\xi^2\vert_{C_2}=2v(v^2+{\subP[S]{u}}^2)^{1/2}+\im{\subP[S]{u}}^2
\label{eq:ixi2onC2}
\end{equation}
is a monotonic function of $v$
and so can be used as a curve parameter
for contour $C_2$
in place of $v$.
If we let $2v(v^2+{\subP[S]{u}}^2)^{1/2}\equiv-\sigma$,
then it follows from
\begin{equation}
\xi\vert_{C_2}=-({\subP[S]{u}}^2+\im\sigma)^{1/2}
\label{eq:xionC2}
\end{equation}
that
\begin{eqnarray}
\int_{C_2}\dif\xi A(\xi,\hat{z})\exp(\im\alpha\xi^2)&=&
    \smfrac{1}{2}\exp[\im(\alpha{\subP[S]{u}}^2-\pi/2)] \nonumber \\
  & & \times\int_0^\infty\dif\sigma\,({\subP[S]{u}}^2+\im\sigma)^{-1/2}
      A\big\vert_{\xi=-({\subP[S]{u}}^2+\im\sigma)^{1/2}}\exp(-\alpha\sigma).
\label{eq:intC2}
\end{eqnarray}
The function $A\vert_{C_2}$
that here enters the integrand
can be determined only by inverting the original transformation (\ref{eq:phi_minus}).

However,
since the dominant contribution
towards the asymptotic value of the above integral
for $\subP{\hat{R}}\gg1$
comes from the vicinity of the boundary point \point{S},
the required inversion of transformation (\ref{eq:phi_minus})
needs to be carried out only to within the leading order in $\sigma$.
The Taylor expansions of $\phi_\pm(\hat{r},\hat{z})$
about $\hat{r}=\subP[S]{\hat{r}}(\hat{z})$
are of the forms
\begin{eqnarray}
\phi_\pm&=& \subP[S]{\phi}+{\subP[S]{\hat{r}}}^{-1}({\subP[S]{\hat{r}}}^2-1)
               ({\subP[S]{\hat{r}}}^2{\subP{\hat r}}^2-1)^{-1/2}(\hat{r}-\subP[S]{\hat{r}}) \nonumber \\
           & & \pm\smfrac{1}{3}(2\subP[S]{\hat{r}})^{3/2}({\subP{\hat r}}^2-1)^{3/2}
               ({\subP[S]{\hat{r}}}^2{\subP{\hat r}}^2-1)^{-3/2}(\hat{r}-\subP[S]{\hat{r}})^{3/2}+\cdots.
\label{eq:Taylorphipm}
\end{eqnarray}
According to Eqs.~\eqref{phi_minus} and \eqref{xiS},
on the other hand,
\begin{equation}
\phi_--\subP[S]{\phi}=\smfrac{1}{2}a(\xi^2-{\subP[S]{\xi}}^2).
\label{eq:phimmphiS}
\end{equation}
In the vicinity of $\xi=\subP[S]{\xi}$, therefore,
Eqs.~\eqref{Taylorphipm} and \eqref{phimmphiS} jointly yield
\begin{equation}
\hat{r}\simeq\subP[S]{\hat{r}}+
  \smfrac{1}{2}\sin^5\subP{\theta}\sec^4\subP{\theta}{\subP{\hat{R}}}^2({\subP[S]{\xi}}^2-\xi^2)
\label{eq:rHatApprox}
\end{equation}
for $\subP{\hat{R}}\gg1$.
Note that ${\subP[S]{\xi}}^2-\xi^2=-\im\sigma$
and that close to the cusp in the far zone
\begin{equation}
\Delta^{1/2}\simeq(2\sin\subP{\theta})^{1/2}\subP{\hat{R}}(\hat{r}-\subP[S]{\hat{r}})^{1/2},
  \qquad\vert\hat{r}-\subP[S]{\hat{r}}\vert\ll1,\qquad\subP{\hat{R}}\gg1.
\label{eq:sqrDeltaApprox}
\end{equation}
Hence,
inserting Eqs.~\eqref{rHatApprox} and \eqref{sqrDeltaApprox} in \Eq{AxizHat},
taking the limit $\subP{\hat{R}}\to\infty$,
and expressing $\xi$ in terms of $\sigma$, 
we find that
\begin{equation}
A\vert_{C_2}\simeq
  \exp[-\im(\mu\subP[C]{\phi}-\pi/4)]{\bf u}_j\subP[S]{\vert}
 \sin\subP{\theta}\sec^2\subP{\theta}({\subP[S]u}^2+\im\sigma)^{1/2}\sigma^{-1/2}
\label{eq:AonC2}
\end{equation}
in the immediate vicinity of the point \point{S} at which $\sigma=0$.

Strictly speaking,
we should excise the singularity of $A$ at $\sigma=0$
by means of an arc-shaped contour.
However, since this singularity is integrable
and so has no associated residue,
the contribution from such a contour vanishes
in the limit that its arc length tends to zero.
An alternative way of handling the removeable singularity at $\sigma=0$,
followed below,
is to introduce a change of integration variable.
If we let $\sigma=\tau^2$,
then the integral in \Eq{intC2} assumes the form
\begin{eqnarray}
\int_{C_2}\dif\xi A(\xi,\hat{z})\exp(\im\alpha\xi^2)&\simeq&
     \sin\subP{\theta}\sec^2\subP{\theta}\exp[-\im(\mu\subP[C]{\phi}+\pi/4)]
     {\bf u}_j\subP[S]{\vert}\int_0^\infty\dif\tau \exp(-\alpha\tau^2) \nonumber \\
 &\simeq&\smfrac{1}{2}(2\pi/\mu)^{1/2}\csc\subP{\theta}\vert\sec\subP{\theta}\vert
     \exp[-\im(\mu\subP[C]{\phi}+\pi/4)]{\bf u}_j\subP[S]{\vert}{\subP{\hat{R}}}^{-1/2},
\label{eq:intC2change}
\end{eqnarray}
where use has been made of Eq.~\eqref{aparam}
and the definition $\alpha\equiv-\mu a/2$.  
Note that this differs from the corresponding expression in Eq.~\eqref{intC1}
for the integral over $C_1$
by the factor $\smfrac{1}{2}\exp(-\im\pi/2)$.

The path of steepest descent
through the boundary point $\xi=\xi_>$,
at which $u=u_>$, $v=0$,
is given by $u=(v^2+{u_>}^2)^{1/2}$,
{\it i.e.}\ by the contour designated as $C_3$ in \Fig{intCont}.
The real part of the exponent
\begin{equation}
\im\xi^2\vert_{C_3}=-2v(v^2+{u_>}^2)^{1/2}+\im{u_>}^2
\label{eq:ixi2onC3}
\end{equation}
is again a monotonic function of $v$
and so can be used to parametrize contour $C_3$ in place of $v$.
If we let $2v(v^2+{u_>}^2)^{1/2}\equiv\chi$,
then it follows from
\begin{equation}
\xi\vert_{C_3}=({u_>}^2+\im\chi)^{1/2}
\label{eq:xionC3}
\end{equation}
that
\begin{eqnarray}
\int_{C_3}\dif\xi A(\xi,\hat{z})\exp(\im\alpha\xi^2)&=&
         \smfrac{1}{2}\exp[\im(\alpha{u_>}^2-\pi/2)] \nonumber \\
& & \times\int_0^\infty\dif\chi\,
     ({u_>}^2+\im\chi)^{-1/2}A\big\vert_{\xi=({u_>}^2+\im\chi)^{1/2}}
     \exp(-\alpha\chi).
\label{eq:intC3}
\end{eqnarray}
The asymptotic value of this integral for $\subP{\hat{R}}\gg1$
receives its dominant contribution from $\chi=0$.
Because the function $A\vert_{C_3}$ is regular,
on the other hand,
its value at $\chi=0$ can be found
by simply evaluating the expression in \Eq{AxizHat} at $\hat{r}=\hat{r}_>$.
The result, for $\subP{\hat{R}}\to\infty$, is
\begin{equation}
A\vert_{C_3,\chi=0}\simeq{\hat{r}_>}^2\sin^4\subP{\theta}\sec^2\subP{\theta}
  ({\hat{r}_>}^2\sin^2\subP{\theta}-1)^{-1}{\bf u}_j\vert_{\hat{r}=\hat{r}_>}
  \exp(-\im\mu\subP[C]{\phi})u_>
\label{eq:AonC3xi0}
\end{equation}
[see Eqs.~\eqref{Delta}) and \eqref{aparam}].
This in conjunction with Watson's lemma
therefore implies that 
\begin{eqnarray}
\int_{C_3}\dif\xi A(\xi,\hat{z})\exp(\im\alpha\xi^2)
  & \simeq &
  2^{1/2}{{\hat r}_>}^2({\hat{r}_>}^2\sin^2\subP{\theta}-1)^{-1}{\bf u}_j\vert_{\hat{r}_>}
\nonumber \\
& & \times\exp[-\im(\mu\phi_-\vert_{\hat{r}_>}+\pi/4)]\mu^{-1}{\subP{\hat{R}}}^{-1},
\label{eq:intC3Approx}
\end{eqnarray}
to within the leading order in ${\subP{\hat{R}}}^{-1}$.

\subsection{Asymptotic value of the radiation field}

The integral in \Eq{intDeltaNonneg}
equals the sum of the three contour integrals
that appear in Eqs.~\eqref{intC1}, \eqref{intC2change} and \eqref{intC3Approx};
the contributions of the contours that connect $C_1$ and $C_2$,
and $C_1$ and $C_3$,
at infinity
(see \Fig{intCont})
are exponentially small
compared to those of $C_1$, $C_2$ and $C_3$ themselves. 
On the other hand,
the leading term in the asymptotic value of the integral over $C_3$
decreases
(with increasing $\subP{\hat{R}}$)
much faster than those of the integrals over $C_1$ and $C_2$:
The integral over $C_3$ decays as ${\subP{\hat{R}}}^{-1}$,
while the integrals over $C_1$ and $C_2$
decay as ${\subP{\hat{R}}}^{-1/2}$
[see Eqs.~\eqref{intC1}, \eqref{intC2change}, and \eqref{intC3Approx}].
The leading term in the asymptotic expansion of the radiation field ${\bf B}^{\rm ns}$
for large $\subP{\hat{R}}$ is therefore given,
according to Eqs.~\eqref{Bns}, \eqref{intDeltaNonneg}, and \eqref{intC1},
by
\begin{eqnarray}
{\bf B}^{\rm ns} & \simeq &
  -\smfrac{2}{3}(1+2\im)(2\pi)^{1/2}{\subP{\hat{R}}}^{-1/2}
  \vert\sec\subP{\theta}\vert\csc\subP{\theta}
  \exp[\im(\Omega/\omega)(\subP{\varphi}+3\pi/2)]
  \sum_{\mu=\mu_\pm}\vert\mu\vert^{1/2} \nonumber \\
& & \times\sgn(\mu)\exp(\im\smfrac{\pi}{4}\sgn\,\mu)
     \sum_{j=1}^3 \bar{q}_j
     \int_{-\infty}^\infty\dif\hat{z}{\bf u}_j
     \subP[C]{\vert}\exp[-\im\mu(\subP[C]{\phi}+\subP{\hat{\varphi}})],
\label{eq:BnsApprox}
\end{eqnarray}
in which $\mu_\pm$ can also be negative
(see the first paragraph of \Section{contours}).

This result agrees with that in Eq.~(55) of Ref.~\citeonline{ArdavanH:Speapc}. 
The two expressions differ by a factor of $2-\im$
because we have here included the additional contribution
that arises from the source elements
in the (vanishingly small) interval $\subP[S]{\hat{r}}\le\hat{r}\le\subP[C]{\hat{r}}$.
The integration with respect to $\hat{r}$
in Eq.~(52) of Ref.~\citeonline{ArdavanH:Speapc}
extends over $\subP[C]{\hat{r}}\le\hat{r}\le\hat{r}_>$,
while that in \Eq{IofzHat}
extends over $\subP[S]{\hat{r}}\le\hat{r}\le\hat{r}_>$.
The contribution from $\subP[S]{\hat{r}}\le\hat{r}\le\subP[C]{\hat{r}}$ is given,
according to Cauchy's theorem,
by the contribution from the lower half of $C_1$
plus the contribution from $C_2$.

Even though the length of the interval $\subP[S]{\hat{r}}\le\hat{r}\le\subP[C]{\hat{r}}$
vanishes as ${\subP{\hat{R}}}^{-2}$
as $\subP{\hat{R}}$ tends to infinity [see \Eq{limxiS}],
the contribution that arises from this interval
towards the value of the field
has the same order of magnitude as that which arises
from $\subP[C]{\hat{r}}\le\hat{r}\le{\hat{r}}_>$,
and is by a factor of order ${\subP{\hat{R}}}^{1/2}$
greater than that which arises from the open interval
$\subP[C]{\hat{r}}<\hat{r}\le{\hat{r}}_>$.
Thus, the nonspherically decaying component of the radiation field
that is observed at any given $(\subP{{\bf x}}, \subP{t})$
arises from those elements of the source,
located at the intersection of the cusp curve
of the bifurcation surface
with the volume of the source (\Fig{cuspProj}),
that occupy the vanishingly small radial interval
\begin{equation}
\delta \hat{r}\equiv\subP[C]{\hat{r}}-\subP[S]{\hat{r}}\simeq
  \smfrac{1}{2}\cos^4\subP{\theta}\csc^5\subP{\theta}{\subP{\hat{R}}}^{-2}
\label{eq:deltar}
\end{equation}
adjacent to the cusp at $\hat{r}=\subP[S]{\hat{r}}\simeq\csc\subP{\theta}$
[see Eqs.~\eqref{rHatC} and \eqref{rHatS}].

The corresponding azimuthal extent of the source
from which the contribution described by \Eq{Bns} arises
is given by the separation $\phi_+-\phi_-$
of the two sheets of the bifurcation surface
shown in \Fig{bifurcation} close to the cusp curve of this surface: 
The contribution of the source elements outside the bifurcation surface
is by a factor of the order of $\subP{\hat{R}}^{-1/2}$ smaller
than those of the elements close to the cusp inside this surface
[see Eqs.~(41) and (42) of Ref.~\citeonline{ArdavanH:Speapc}].
Since $\phi_+-\phi_-\simeq
(2^{5/2}/3)(\csc\subP{\theta})^{-3/2}(\hat{r}-\subP[S]{\hat{r}})^{3/2}$
for $\vert\hat{r}-\subP[S]{\hat{r}}\vert\ll1$ and $\subP{\hat{R}}\gg1$
[see \Eq{Taylorphipm}],
and the contributing interval in $\hat{r}$
is of the order of ${\subP{\hat{R}}}^{-2}$
[see \Eq{deltar}],
it follows that the contributing interval in $\varphi$ is 
\begin{equation}
\delta\varphi\equiv(\phi_+-\phi_-)_{\hat{r}
  =\subP[C]{\hat{r}}}\simeq\smfrac{2}{3}\cot^6\subP{\theta}{\subP{\hat{R}}}^{-3}.
\label{eq:deltavarphi}
\end{equation}
The contribution
from this vanishingly small azimuthal extent of the rotating source
is made when the retarded position of this part of the source
is $\varphi=\subP[C]{\varphi}$
[see \Eq{varphiC}],
{\it i.e.}\ when the contributing source elements
approach the observer with the speed of light and zero acceleration
along the radiation direction.
Thus,
the source that generates the nonspherically decaying field
observed at a point $(\subP{\hat{R}},\subP{\theta},\subP{\varphi})$
in the far zone ($\subP{\hat{R}}\gg1$)
consists entirely of the narrow filament parallel to the $z$ axis
that occupies a radial interval $\delta\hat{r}\propto{\subP{\hat{R}}}^{-2}$
encompassing $\hat{r}\simeq\csc\subP{\theta}$
and an azimuthal interval $\delta\varphi\propto{\subP{\hat{R}}}^{-3}$
encompassing $\varphi\simeq\subP{\varphi}+3\pi/2$ at the retarded time.
\subsection{Frequency independence of the nonspherical decay%
\label{sec:frequency}}
A further implication of the above analysis
is that the generated field decays nonspherically
irrespective of what the values of the frequencies $\mu_\pm\omega$ may be.
There is no approximation involved
in introducing the transformation (\ref{eq:phi_minus}),
and the asymptotic expansion is for large $\alpha$.
As derived here, therefore,
the only condition for the validity of \Eq{BnsApprox}
is that the absolute value of
$\alpha\simeq\frac{1}{2}\mu_\pm\subP{\hat{R}}\sin^4\subP{\theta}\sec^2\subP{\theta}$
should be large,
a condition that is automatically satisfied in the far zone
for all nonzero frequencies.

That the leading term
in the asymptotic expansion of the integral in \Eq{intDeltaNonneg}
is proportional to ${\subP{\hat{R}}}^{-1/2}$,
instead of ${\subP{\hat{R}}}^{-1}$,
is a consequence of the particular features
of the phase function $\phi_-$
described by Eqs.~\eqref{phi_minusOnrHatC}--\eqref{aparam}.
These features originate in and reflect
the particular properties of the time-domain phase $t+R(t)/c$;
they are totally independent
of both the wavelength of the radiation
and the size of the source.
In contrast
to all other nonspherically decaying solutions of Maxwell's equations
reported in the published literature
(see, {\it e.g.}, Refs.~\citeonline{ShaarawiAM:Genafp,ReiveltKaido:Expdro,RecamiE:thelss}),
whose slow spreading and decay
only occur within the Fresnel distance from the source,
the nonspherical decay that is discussed here
remains in force at all distances. 
In fact,
the greater the distance $\subP{R}$ from the source,
the more the leading term dominates the asymptotic approximation in \Eq{BnsApprox}.

The remaining $\hat{z}$ integration
in the above expression for ${\bf B}^{\rm ns}$
amounts to a Fourier decomposition
of the source densities $s_{r,\varphi,z}\subP[C]{\vert}$
with respect to $\hat{z}$.
Using Eqs.~\eqref{phi_minusOnrHatC}--\eqref{RhatC}
to replace $\subP[C]{\phi}$ in \Eq{BnsApprox}
by its far-field value
\begin{equation}
\subP[C]{\phi}\simeq\subP{\hat{R}}-\hat{z}\cos\subP{\theta}+3\pi/2,
\label{eq:phiCapprox}
\end{equation}
and using \Eq{u_j}
to write out ${\bf u}_j$
in terms of $s_{r,\varphi,z}$,
we find that the electric field
${\bf E}^{\rm ns}={\hat{\bf n}}{\bf\times B}^{\rm ns}$
of the nonspherically decaying radiation is given by
\begin{eqnarray}
{\bf E}^{\rm ns}&\simeq&
    \smfrac{4}{3}(2\pi)^{1/2}{\subP{\hat{R}}}^{-1/2}\vert\sec\subP{\theta}\vert\csc\subP{\theta}
    \exp[\im(\Omega/\omega)(\subP{\varphi}+3\pi/2)]
    \sum_{\mu=\mu_\pm}\vert\mu\vert^{1/2} \nonumber \\
& &\times\sgn(\mu)\exp(\im\smfrac{\pi}{4}\sgn\,\mu)
    \exp[-\im\mu(\subP{\hat{R}}+\subP{\hat{\varphi}}+3\pi/2)]
    \{(\im\bar{s}_\varphi+\Omega\bar{s}_r/\omega){\hat{\bf e}}_\parallel \nonumber \\
& &-[(\im\bar{s}_r-\Omega\bar{s}_\varphi/\omega)\cos\subP{\theta}
    +\Omega\bar{s}_z\sin\subP{\theta}/\omega]{\hat{\bf e}}_\perp\},
\label{eq:EnsApprox}
\end{eqnarray}
in which $\bar{s}_{r,\varphi,z}$
stand for the following Fourier transforms
of $s_{r,\varphi,z}\subP[C]{\vert}$
with respect to $\hat{z}$:
\begin{equation}
\bar{s}_{r,\varphi,z}\equiv
  \int_{-\infty}^\infty\dif\hat{z}\,
  s_{r,\varphi,z}(\hat{r},\hat{z})\big\vert_{\hat{r}=\csc\subP{\theta}}
  \exp(\im\mu\hat{z}\cos\subP{\theta}).
\label{eq:sbar}
\end{equation}
This field is observable
only at those polar angles $\subP{\theta}$
within the interval
$\arccos(1/\hat{r}_<)\le\vert\subP{\theta}-\pi/2\vert\le\arccos(1/\hat{r}_>)$
for which $s_{r,\varphi,z}\vert_{\hat{r}=\csc\subP{\theta}}$ are nonzero,
{\it i.e.} at those observation points
(outside the plane of rotation)
the cusp curve of whose bifurcation surface (\Fig{bifurcation})
intersects the source distribution (\Fig{cuspProj}).
\subsection{Relevance to computational models of the emission from superluminal sources}
The asymptotic analysis outlined in this section
provides a basis also
for the computational treatment
of the nonspherically decaying field ${\bf B}^{\rm ns}$.
The original formulation of ${\bf B}^{\rm ns}$
appearing in \Eq{Bns},
in which the integral has a rapidly oscillating kernel,
is not suitable for computing a field
whose value in the radiation zone
receives its main contribution
from such small fractions of the $\hat{r}$ and $\hat{\varphi}$ integration domains
as $\delta\hat{r}\propto{\subP{\hat{R}}}^{-2}$
and $\delta\varphi\propto{\subP{\hat{R}}}^{-3}$. 
The above conversion of the Fourier-type integrals
to Laplace-type integrals
renders the selecting out and handling
of the contributions from integrands
with such narrow supports
numerically more feasible.

\section{The collection of nondiffracting subbeams 
delineating the overall distribution of the nonspherically decaying radiation%
\label{sec:ndsubbeams}}

We have seen that the wave fronts
that emanate from a given volume element
of a rotating superluminal source
possess an envelope
consisting of two sheets
that meet along a cusp (\Fig{envelope}).
There is a higher-order focusing
involved in the generation of the cusp
than in that of the envelope itself,
so that the intensity of the radiation
from an extended source
attains its maximum
on the bundle of cusps
that are emitted by various source elements.
If a source element approaches an observeration point \point{P}
with the speed of light and zero acceleration along the radiation direction,
then the cusp it generates passes through \point{P}.
The reason is
that both the locus of source elements
that approach the observer with the speed of light and zero acceleration
[{\it i.e.}\ the cusp curve of the bifurcation surface (\Fig{bifurcation})]
and the cusp that is generated
by a given source element
are described by the same equation:
The cusp curve of the bifurcation surface
resides in the space of source points
and so is given by \Eq{cuspcurve}
for a fixed $(\subP{\hat r},\subP{\varphi},\subP{\hat{z}})$,
while the cusp curve of the envelope
resides in the space of observation points
and is given by \Eq{cuspcurve}
for a fixed $(r,\varphi,z)$.
\subsection{Angular extent of the nonspherically decaying emission}
The collection of cusp curves
that are generated by the constituent volume elements of an extended source
thus defines what might loosely be termed a `radiation beam',
although its characteristics are distinct from those of conventionally produced beams.
The field decays nonspherically
only along the bundle of cusp curves embodying this radiation beam.
Since the cusp that is generated
by a source element with the radial coordinate $\hat{r}$
lies on the cone $\subP{\theta}=\arcsin(1/\hat{r})$ in the far zone,
the nonspherically decaying radiation
that arises from a source distribution
with the radial extent $\hat{r}_<\le\hat{r}\le\hat{r}_>$
is detectable only within the conical shell
$\arcsin(1/\hat{r}_>)\le\subP{\theta}\le\arcsin(1/\hat{r}_<)$.

The field that is detected at a given point \point{P}
within this conical shell
arises almost exclusively from a filamentary part of the source
parallel to the $z$ axis
whose radial and azimuthal extents
are of the order of $\delta \hat{r}\propto{\subP{\hat{R}}}^{-2}$
and $\delta\varphi\propto{\subP{\hat{R}}}^{-3}$,
respectively
[see Eqs.~\eqref{deltar} and \eqref{deltavarphi}].
The bundle of cusps
emanating from this narrow filament
occupies a much smaller solid angle
than that described above.
The parametric equation $\subP{\hat{z}}=\subP{\hat{z}}(\subP{\hat r})$,
$\subP{\varphi}=\subP{\varphi}(\subP{\hat r})$
of the particular cusp curve
that emanates from a given source element
$(r,\hat{\varphi},z)$ can be written, using \Eq{cuspcurve}, as
\begin{equation}
\subP{\hat{z}}=\hat{z}\pm({\subP{\hat r}}^2-1)^{1/2}(\hat{r}^2-1)^{1/2},
\label{eq:cuspElemzHat}
\end{equation}
\begin{equation}
\subP{\varphi}=\varphi-2\pi+\arccos[1/(\hat{r}\subP{\hat r})].
\label{eq:cuspElemvarphi}
\end{equation}
If we rewrite Eqs.~\eqref{cuspElemzHat} and \eqref{cuspElemvarphi}
in terms of the spherical polar coordinates $(\subP{R},\subP{\theta},\subP{\varphi})$
of the observation point \point{P}
and solve them for $\subP{\theta}$ and $\subP{\varphi}$
as functions of $(r,z)$ and $\subP{R}$,
we find that the cusp that is generated
by a source point with the coordinates $(r,\hat{\varphi},z)$
passes through the following two points
on a sphere of radius $\subP{R}$: 
\begin{equation}
\subP{\theta}=\arccos\left\{
    \frac{1}{\hat{r}\subP{\hat{R}}}\left[\frac{\hat{z}}{\hat{r}}\pm(\hat{r}^2-1)^{1/2}
      \left(\subP{\hat{R}}^2-1-\frac{\hat{z}^2}{\hat{r}^2}\right)^{1/2}\right]\right\},
\label{eq:cuspRpinter1}
\end{equation}
and
\begin{equation}
\subP{\varphi}=\varphi-2\pi+\arccos[1/(\subP{\hat{R}}\hat{r}\sin\subP{\theta})],
\label{eq:cuspRpinter2}
\end{equation}
where the $\pm$
correspond to the two halves of this cusp curve
above and below the plane of rotation
(see \Fig{cuspProj}).

The Taylor expansion
of the expressions on the right-hand sides of Eqs.~\eqref{cuspRpinter1} and \eqref{cuspRpinter2}
in powers of ${\subP{\hat{R}}}^{-1}$
yields
\begin{equation}
\subP{\theta}=\arcsin(1/\hat{r})
  -(\hat{z}/\hat{r}){\subP{\hat{R}}}^{-1}
  \pm\smfrac{1}{2}(\hat{r}^2-1)^{1/2}{\subP{\hat{R}}}^{-2}+\cdots,
\label{eq:cuspRpinter1Taylor}
\end{equation}
and
\begin{equation}
\subP{\varphi}=\varphi-3\pi/2-{\subP{\hat{R}}}^{-1}+\cdots.
\label{eq:cuspRpinter2Taylor}
\end{equation}
These show
that incremental changes $\delta r$, $\delta z$ and $\delta\varphi$
in the position of the source element $(r,\hat{\varphi},z)$
result in the following changes
in the $(\subP{\theta},\subP{\varphi})$ coordinates
of the point at which the cusp arising from that element
intersects a sphere of radius $\subP{R}$ in the far zone:
\begin{equation}
\delta\subP{\theta}
  =-\hat{r}^{-1}(\hat{r}^2-1)^{-1/2}\delta\hat{r}-\hat{r}^{-1}\delta\hat{z}{\subP{\hat{R}}}^{-1}
  +\cdots,
\label{eq:deltathetap}
\end{equation}
and $\delta\subP{\varphi}=\delta\varphi+\cdots$. 
Because $\delta \hat{r}$ is of the order of ${\subP{\hat{R}}}^{-2}$,
while $\delta\hat{z}$ is of the order of unity
for the filamentary source of the field that is detected at \point{P},
the dominant term
in the expression on the right-hand side of \Eq{deltathetap}
is that proportional to the $\hat{z}$ extent of the filament.
Given the observation point \point{P},
and hence a set of fixed values
for the dimensions ($\delta \hat{r}$, $\delta\varphi$, $\delta\hat{z}$)
of the filamentary source
of the field that is detected at \point{P},
it therefore follows that the bundle of cusps generated by such a filament
occupies a solid angle with the dimensions 
\begin{equation}
\delta\subP{\theta}\simeq-\delta\hat{z}\sin\subP{\theta}{\subP{\hat{R}}}^{-1}
  \quad{\rm and}\quad\delta\subP{\varphi}\simeq\delta\varphi
\label{eq:deltathetapApprox}
\end{equation}
in the far zone.
[Here, we have made use of the fact
that $\subP{\theta}\simeq\arcsin(1/\hat{r})$
to within the zeroth order in ${\subP{\hat{R}}}^{-1}$
to express $\hat{r}$ in \Eq{deltathetapApprox}
in terms of $\subP{\theta}$.]

The bundle of cusp curves
occupying the solid angle \eqref{deltathetapApprox}
embodies a subbeam
that does not diverge in the direction of $\subP{\theta}$.
The polar width $\delta\subP{\theta}$ of this subbeam
decreases with the distance $\subP{\hat{R}}$
in such a way that the thickness
$\subP{\hat{R}}\delta\subP{\theta}$ of the subbeam
in the polar direction
remains constant:
It equals the projection of the $\hat{z}$ extent, $\delta\hat{z}$,
of the contributing filamentary source
onto a direction normal to the line of sight
at all $\subP{\hat{R}}$.
The azimuthal width of the subbeam,
on the other hand,
diverges as does any other radiation beam:
$\delta\subP{\varphi}$ is independent of $\subP{\hat{R}}$
(see \Fig{subbeam}).

Thus,
the bundle of cusps
that emanates from the filamentary locus
of the set of source elements
responsible for the nonspherically decaying field at \point{P}
intersects a large sphere of radius $\subP{R}$
(enclosing the source)
along a strip
the thickness of whose narrow side
is independent of $\subP{R}$.
According to \Eq{deltathetapApprox},
the area $\subP{R}^2\sin\subP{\theta}\delta\subP{\theta}\delta\subP{\varphi}$
subtended by this subbeam
increases as $\subP{R}$,
rather than $\subP{R}^2$,
with the radius of the sphere enclosing the source.
Conservation of energy demands, therefore,
that the Poynting vector associated with this radiation
should correspondingly decrease as $\subP{R}^{-1}$
instead of $\subP{R}^{-2}$,
in order that the flux of energy remain the same
across various cross sections of the subbeam.
This requirement is,
of course,
automatically met
by the (nonspherical) rate of decay
of the intensity of the radiation
that propagates along the subbeam.

The nondiffracting subbeams
that are detected at two distinct observation points
within the solid angle $\arcsin(1/\hat{r}_>)\le\subP{\theta}\le\arcsin(1/\hat{r}_<)$
arise from two distinct filamentary parts
of the source with essentially no common elements
[see Eqs.~\eqref{deltar} and \eqref{deltavarphi}].
The subbeam that passes through an observation point ${\rm P}^\prime$,
though sharing the same general properties as that which passes through \point{P},
arises from those elements of the source,
located at $\hat{r}^\prime=\csc\subP[P^\prime]{\theta}$,
$\varphi^\prime=\subP[P^\prime]{\varphi}$,
that approach ${\rm P}^\prime$,
rather than \point{P},
with the speed of light and zero acceleration
at the retarded time.
Not only are the focused wave packets that embody the cusp
constantly dispersed and reconstructed
out of other (spherically spreading) waves,\cite{ArdavanH:Genfnd}
but also the filaments
that act as sources of these focused waves
each occupy a vanishingly small
($\sim{\subP{\hat{R}}}^{-5}$)
disjoint volume of the overall source distribution,
and so are essentially independent of one another.
Unlike conventional radiation beams,
 which have fixed sources,
the subbeam that passes through an observation point \point{P}
arises from a source whose location and extent depend on \point{P}.

It would be possible to identify the individual nondiffracting subbeams
only in the case of a source
whose length scale of spatial variations
is comparable to ${\subP{\hat{R}}}^{-2}$
({\it e.g.}\ in the case of a turbulent plasma
with a superluminally rotating macroscopic distribution).
The overall beam
within which the nonspherically decaying radiation is detectable
would then consist of an incoherent superposition
of coherent, nondiffracting subbeams
with widely differing amplitudes and phases.
The individual coherent subbeams
decay nonspherically,
but the incoherence of their phase relationships
ensures that the integrated flux of energy
associated with their superposition across this finite solid angle
remains independent of $\subP{R}$.
Note that the individual subbeams constituting the overall beam
would be narrower and more distinguishable,
the farther the observer is from the source.

\section{Concluding remarks\label{sec:concluding}}

The analysis we have presented here
was motivated by questions
encountered in the course of the design, construction, and testing
of practical machines
for investigating the emission from superluminal sources.\cite{ArdavanA:Exponr}
The original mathematical treatment
of the nonspherically decaying radiation,\cite{ArdavanH:Speapc}
in which the integral over the volume of the source
has a rapidly oscillating kernel,
is not suitable
for the computational modeling of the emission from such machines.
We have seen that the nonspherically decaying radiation
detected in the radiation zone
receives its main contribution
from such small fractions
of the radial and azimuthal integration domains
as $\delta\hat{r}\sim{\subP{\hat{R}}}^{-2}$ and $\delta\varphi\sim{\subP{\hat{R}}}^{-3}$.
The above conversion of the Fourier-type integrals
to Laplace-type integrals
renders the selecting out and handling
of the contributions from integrands with such narrow supports
numerically more feasible.

Not only the nonspherical decay of its intensity,
but also the narrowness of both the beam into which it propagates
and the region of the source from which it arises
are features that are unique to the emission
from a rotating superluminal source.
These features are not shared by any other known emission mechanism.
On the other hand,
they are remarkably similar to the observed features 
of an emission that has long been known to radio astronomers:
to the (as yet unexplained) extreme properties
of the giant pulses that are received from pulsars
(see, {\it e.g.} Refs.~%
\citeonline{LorimerD:HbPA,SallmenS:Simdog,HankinsTH:Nanrbs,SoglasnovVA:GiapPB,PopovMV:GPmcrecp}).
The giant radio pulses from the Crab pulsar
have a temporal structure of the order of a nanosecond.\cite{HankinsTH:Nanrbs}
Under the assumption that they decay spherically
like other conventional emissions,
the observed values of these pulses' fluxes
imply that their energy densities
generally exceed the energy densities
of both the magnetic field and the plasma
in the magnetosphere of a pulsar
by many orders of magnitude.\cite{SoglasnovVA:GiapPB}
``The plasma structures responsible for these emissions
must be smaller than one metre in size,
making them by far the smallest objects
ever detected and resolved outside the Solar System,
and the brightest transient radio sources in the sky.''\cite{HankinsTH:Nanrbs}

The highly stable periodicity
of the mean profiles of the observed pulses,\cite{LorimerD:HbPA}
{\it i.e.}\ the rigidly rotating distribution of the radiation from pulsars,
can only arise from a source
whose distribution pattern correspondingly rotates rigidly,
a source whose average density depends on the azimuthal angle $\varphi$
in the combination $\varphi-\omega t$ only:
Maxwell's equations demand
that the charge and current densities
that give rise to this radiation
should have exactly the same symmetry
($\partial/\partial t=-\omega\partial/\partial\varphi$)
as that of the observed radiation fields
${\bf E}$ and ${\bf B}$.
On the other hand,
the domain of applicability of such a symmetry casnnot be localized;
a solution of Maxwell's equations that satisfies this symmetry
applies either to the entire magnetosphere
or to a region whose boundary is an expanding wave front.
Unless there is no plasma outside the light cylinder, therefore,
the macroscopic distribution of electric current in the magnetosphere of a pulsar
should have a superlumially rotating pattern in $r>c/\omega$.
The superluminal source described by \Eq{elecPol}
captures the essential features of the macroscopic charge-current distribution
that is present in the magnetosphere of a pulsar
and is thus an inevitable implication
of the observational data on these objects.

Once it is acknowledged
that the source of the observed giant pulses
should have a superluminally rotating distribution pattern,
the extreme values of their brightness temperature ($\sim10^{39}$ $^\circ$K),
temporal width ($\sim1$ ns),
and source dimension ($\sim1$ m)
are all explained by the results of the above analysis.
The nonspherical decay of the resulting radiation
would imply that the energy density
and so the brightness temperature
of the observed pulses
are by a factor of the order of $\hat{R}/(\hat{r}_>-\hat{r}_<)^2$
smaller than those that are normally estimated
by using an inverse-square law,\cite{ArdavanH:Speapc}
a factor that ranges from $10^{15}$ to $10^{25}$
in the case of known pulsars.\cite{LorimerD:HbPA}

The nondiffracting nature
of this nonspherically decaying radiation [\Eq{deltathetapApprox}], 
together with its arising only
from the filamentary part of the source
that approaches the observer with the speed of light and zero acceleration
[Eqs.~\eqref{deltar} and \eqref{deltavarphi}],
likewise explain the values of its temporal width and source dimension.
Furthermore,
that the overall beam
within which the nonspherically decaying radiation is detectable
should in general consist of an incoherent superposition
of coherent, nondiffracting subbeams (\Section{ndsubbeams})
is consistent with the conclusion reached by Popov et al.\cite{PopovMV:GPmcrecp}
that
``the radio emission of the Crab pulsar
at the longitudes of the main pulse and interpulse
consists entirely of giant pulses.''\cite{PopovMV:GPmcrecp}

Two other features
of the emission from a rotating superluminal source
that were derived elsewhere\cite{SchmidtA:Occopmlw,ArdavanH:Fresfb}
are also consistent with the observational data on pulsars:\cite{LorimerD:HbPA}
the occurrence of concurrent `orthogonal' polarization modes
with swinging position angles\cite{SchmidtA:Occopmlw}
and a broadband frequency spectrum.\cite{ArdavanH:Fresfb}

\section*{Acknowledgment}
This work is supported by U.\ S.\ Department of Energy grant LDRD 20050540ER.








\bibliographystyle{osajnl}
\bibliography{jabosa,superluminal}






\end{document}